# Classification, potential role, and modeling of power-to-heat and thermal energy storage in energy systems: A review

Md. Nasimul Islam Maruf [a,b,*], Germán Morales-España [c], Jos Sijm [c], Niina Helistö [d], Juha Kiviluoma [d]

[a] Department of Energy and Environmental Management, Europa-Universität Flensburg, Germany
[b] Department of Electrical Energy Storage, Fraunhofer Institute for Solar Energy Systems ISE, Germany
[c] TNO Energy Transition, Amsterdam, The Netherlands
[d] VTT Technical Research Centre of Finland, Espoo, Finland

ABSTRACT

Most of the power-to-heat and thermal energy storage technologies are mature and impact the European energy transition. However, detailed models of these technologies are usually very complex, making it challenging to implement them in large-scale energy models, where simplicity, e.g., linearity and appropriate accuracy, are desirable due to computational limitations. In the literature, the main power-to-heat and thermal energy storage technologies across all sectors have not been clearly identified and characterized. Their potential roles have not been fully discussed from the European perspective, and their mathematical modeling equations have not been presented in a compiled form. This paper contributes to the research gap in three main parts. First, it identifies and classifies the major power-to-heat and thermal energy storage technologies that are climate-neutral, efficient, and technologically matured to supplement or substitute the current fossil fuel-based heating. The second part presents the technology readiness levels of the identified technologies and discusses their potential role in a sustainable European energy system. The third part presents the mathematical modeling equations for the technologies in large-scale optimization energy models. We identified electric heat pumps, electric boilers, electric resistance heaters, and hybrid heating systems as the most promising power-to-heat options. We grouped the most promising thermal energy storage technologies under four major categories. Low-temperature electric heat pumps, electric boilers, electric resistance heaters, and sensible and latent heat storage show high technology readiness levels to facilitate a large share of the heat demand. Finally, the mathematical formulations capture the main effects of the identified technologies.

Upper-case letters are used for denoting parameters and sets. Lower-case letters denote variables and indexes.

## Introduction

### *Background*

Sustainable heating and cooling strategies are expected to play a significant role in achieving the ambitious Paris Agreement target in Europe. Heating and cooling in buildings and industries account for half of the total final energy demand in the European Union (EU), of which 80 % is industrial process heating [1]. In the EU households, space and water heating account for 79 % of the total final energy use [2,3]. Over the years 2003–2018, the share of energy from renewable sources for heating and cooling in these two sectors has steadily grown from 10 % to almost 20 % in the EU-27 and the United Kingdom (UK) [4]. Fig. 1 shows

*Abbreviations:* °C, Degree Celsius; CAES, Compressed Air Energy Storage; CHP, Combined Heat and Power; $CO_2$, Carbon dioxide; COP, Coefficient of Performance; ETS, Electrical Thermal Storage; EU, European Union; FOR, Feasible Operating Region; GWh, Gigawatt-hour; HRE, Heat Roadmap Europe; HTHP, High-temperature Heat Pump; HTPCM, High-temperature PCM; K, Kelvin; kV, Kilovolt; LAES, Liquid Air Energy Storage; LHS, Latent Heat Storage; LP, Linear Programming; LPCM, Low-temperature PCM; MILP, Mixed-integer Linear Programming; MW, Megawatt; NLP, Nonlinear Programming; P2H, Power-to-heat; PCM, Phase Change Material; PHS, Passive Heat Storage; PJ, Petajoule; SHS, Sensible Heat Storage; SPCM, Sub-zero Temperature PCM; SSTS, Solid State Thermal Storage; TCS, Trade, Commerce, and Services; TES, Thermal Energy Storage; THS, Thermo-chemical Heat Storage; TMS, Thermo-mechanical Heat Storage; TRL, Technology Readiness Level; TTES, Tank Thermal Energy Storage; TWh, Terawatt-hour; UK, United Kingdom; UTES, Underground Thermal Energy Storage; V, Volt; VRE, Variable Renewable Energy.

* Corresponding author at: Department of Energy and Environmental Management, Europa-Universität Flensburg, Germany.
*E-mail addresses:* ni.maruf@uni-flensburg.de, md.nasimul.islam.maruf@ise.fraunhofer.de (Md.N.I. Maruf).

**Nomenclature**

*Indexes and Sets*

| | |
|---|---|
| $j$ | Index of extreme point |
| $J$ | Set of extreme points of CHP |
| $t$ | Index of time steps |
| $T$ | Set of time steps |

*Variables*

| | |
|---|---|
| $c_{CHP}$ | Operational cost of the convex CHP unit |
| $c_{CHP}^{SU}$ | Startup cost of the CHP unit |
| $p_{CHP}$ | Power production of the CHP unit |
| $p_{EB}$ | Electrical power input of the electric boiler |
| $p_{HB}$ | Electrical power input of the hybrid heating system |
| $p_{HP}$ | Electrical power input of the heat pump |
| $p_{SUP}$ | Electrical power input of the supplementary equipment |
| $\dot{q}_C$ | Heat flow to the storage unit (charging) |
| $q_{CHP}$ | Heat production of the CHP unit |
| $\dot{q}_D$ | Heat flow from the storage unit (discharging) |
| $\dot{q}_{EB}$ | Output heat flow of the electric boiler |
| $\dot{q}_{GB}$ | Output heat flow of the gas boiler |
| $\dot{q}_{HP}$ | Output heat flow of the heat pump |
| $\dot{q}_{SUP}$ | Output heat flow of the supplementary equipment |
| $s$ | Heat storage level |
| $u$ | Binary variable which is equal to 1 if the unit is producing above minimum output and 0 otherwise. |
| $x_j$ | Variable used to encode the convex combination of the operating region of the CHP unit |

*Parameters*

| | |
|---|---|
| $COP$ | Actual (real) coefficient of performance of the heat pump |
| $C_j$ | Production cost of characteristic point $j$ of the CHP unit |
| $COP_{Theory}$ | Theoretical maximum coefficient of performance of the heat pump |
| $C_{CHP}^{SU}$ | Maximum startup cost of CHP unit |
| $COP_{Theory}^{Cooling}$ | Theoretical maximum coefficient of performance of the heat pump for cooling |
| $L_D$ | Dynamic heat loss of the heat storage unit |
| $L_s$ | Stationary heat loss of the heat storage unit |
| $\underline{P_{CHP}}$ | Minimum power production of the CHP unit |
| $\overline{P_{CHP}}$ | Maximum power production of the CHP unit |
| $\underline{P_{EB}}$ | Minimum power input of the electric boiler |
| $\overline{P_{EB}}$ | Maximum power input of the electric boiler |
| $\overline{P_{HP}}$ | Maximum power consumption of the CHP |
| $P_j$ | Power production of characteristic point $j$ of the CHP unit |
| $\dot{Q}$ | Total heat demand |
| $\overline{Q_{CHP}}$ | Maximum heat production of the CHP unit |
| $\overline{Q_{EB}}$ | Maximum heat output of the electric boiler |
| $\overline{Q_{FLOW}}$ | Maximum allowable heat flow from and to the storage unit |
| $Q_H$ | Heat supplied to the high temperature reservoir |
| $Q_j$ | Heat production of characteristic point $j$ of the CHP unit |
| $Q_L$ | Heat extracted from the low temperature reservoir |
| $RD$ | Ramp down rate of the CHP unit |
| $RU$ | Ramp up rate of the CHP unit |
| $S_{cap}$ | Capacity of the heat storage unit |
| $T_H$ | Temperature of the high reservoir |
| $T_L$ | Temperature of the low reservoir |
| $\beta$ | Power loss index (ratio between lost power generation and increased heating generation) |
| $\Delta T$ | Temperature difference between high temperature and low temperature reservoirs |
| $\eta_{EB}$ | Efficiency of the electric boiler |
| $\eta_{HP}$ | Efficiency of the heat pump |
| $\sigma$ | Power-to-heat ratio of the CHP unit |

the share of renewable energy as a percentage of gross final energy consumption for heating and cooling in European countries in 2018. In Sweden, Latvia, Finland, and Estonia, renewable energy accounted for more than 50 % of the energy consumption for heating and cooling. On the other hand, the shares are less than 10 % in Ireland, the Netherlands, Belgium, and Luxembourg.

Power-to-heat technologies, often abbreviated as PtH or P2H, refer to applications in which electrical energy generates heat, which is mainly used in the built environment or industrial processes. P2H offers many advantages to drive the energy transition. For example, P2H using excess variable renewable energy (VRE) helps energy regulation and reduces the use of fossil fuels. With more VRE penetration on the European grid, the power system's voltage, transient, small-signal, and frequency stabilities are increasingly challenged [5]. For this reason, grid operators occasionally switch off some of the (inverter-based) VRE plants (e.g., wind turbines), leading to the curtailment of a large amount of renewable energy. Using P2H, the total load can be increased during high VRE production and low load, thus limiting the rise of the instantaneous VRE penetration and keeping the power system more stable with less curtailment of renewable energy. Besides, P2H offers additional flexibility in the electricity market by using and balancing energy in times of low or negative electricity prices. When there is a negative electricity price, fossil-fueled power plants may not leave the market because it might not be profitable, e.g., they reserve capacity for the balancing energy market. P2H can provide this service at a comparatively low cost and thereby reduce carbon dioxide emissions. Efficient P2H systems can ensure the energy supply system's stability and contribute to the decarbonization of the heating sector using green electricity, and thus actively supporting the energy transition.

A heat pump is an efficient P2H application that can extract and provide heat from a medium (water or air) with much less electrical energy use, i.e., one electricity energy unit of input typically produces more than one or two heat energy units of output. Another typical application of P2H is domestic hot water production by using electric boilers or instantaneous water heaters. P2H applications also include room heating using direct electric heaters such as storage heaters or radiant heaters. For the central provision of large amounts of heating energy, auxiliary P2H applications sometimes support district heating grids. Hybrid systems are becoming increasingly popular for increased flexibility in large-scale power systems, where the P2H technologies integrate with other heating technologies or a combined heat and power (CHP) plant. The integration of heat storage tanks in hybrid systems further enhances flexibility. Because of its high efficiency and comparatively low cost, P2H from renewable sources is a matter of growing interest across Europe. Studies and projects on P2H present potentials and opportunities for applying P2H in different sectors of the EU energy system [6,7]. P2H applications, especially in hybrid systems or combined with heat storage, support VRE integration in the energy system in the following ways [8]:

1. Reduce VRE curtailments;
2. Increase demand-side flexibility through load shifting;
3. Provide grid services via aggregators to optimize heating costs for consumers and provide grid balancing services to the national grid; and
4. Increase self-consumption from local renewable-based generation.

Heat pumps, electric boilers, and electric resistance heaters are generally identified as the three most promising P2H applications. A combination of two or more heating options is also possible, typically presented as a hybrid heating system.

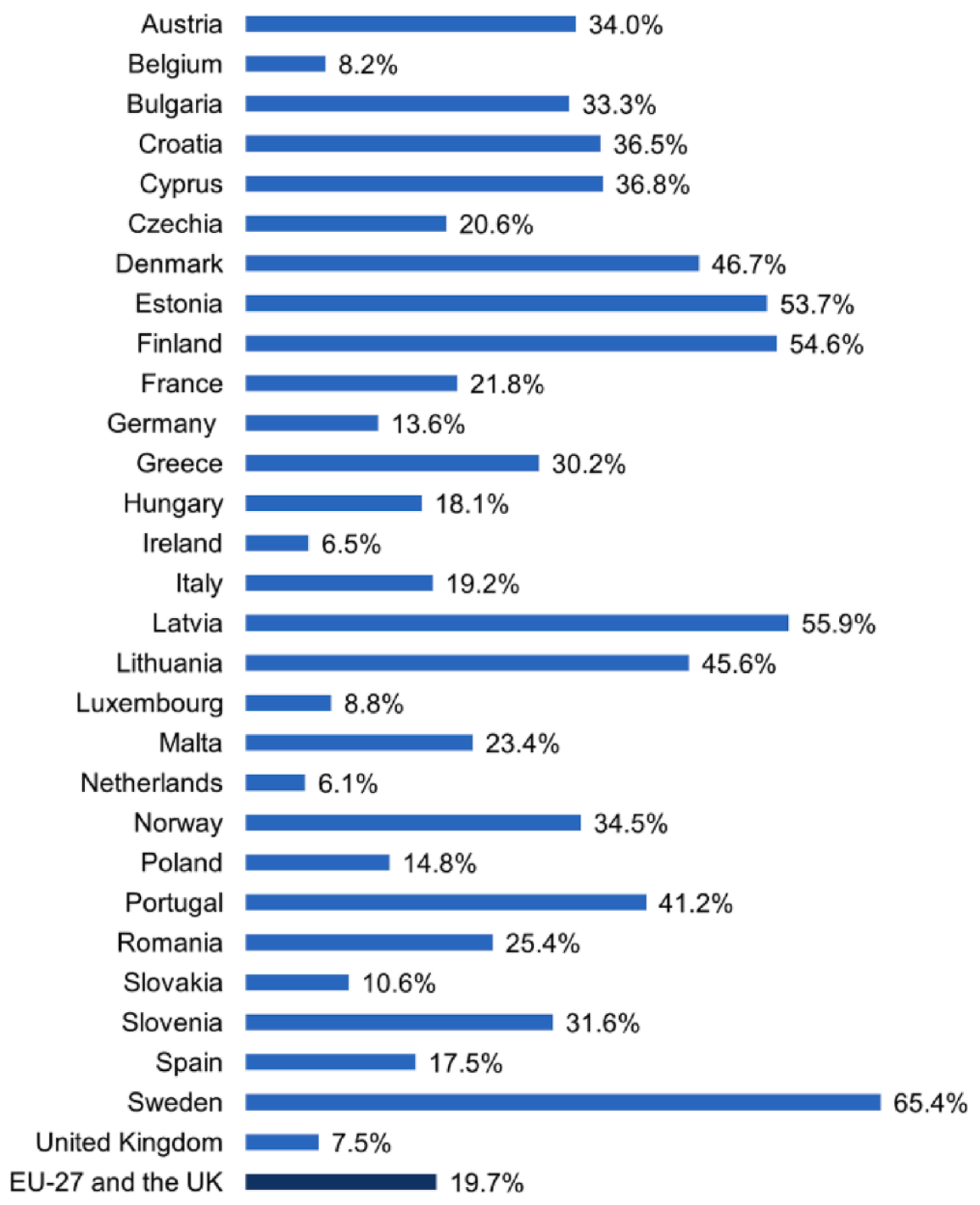

**Fig. 1.** Share of renewable energy for heating and cooling in European countries in 2018. Data Source: Eurostat [4].

Heat pump is an outstanding technology to provide flexibility to the power system while providing efficient heating and cooling solutions. Using external energy, they use a refrigerant through insulated pipes to transfer heat from a low-temperature source to a high-temperature sink. Heat pumps are used for space and water heating, air conditioning, and diverse industrial applications. The technology is advantageous because of its high efficiency, low energy cost, easy installation, minimum maintenance requirement, and high safety standards compared to other heating technologies.

Electric boiler is another popular P2H application widely used in utility-related processes to produce hot water and steam. The technology is advantageous because of its low initial cost, robust and compact design, flexibility, no stack requirements, zero carbon emission if provided with renewable electricity, quiet operation, high efficiency, and ease of maintenance. Two installations are typical: electrical resistance boiler and electrode boiler. In addition, there are also small-scale infrared and induction-type electric boilers. While electric resistance boilers are connected at low voltages, electrode boilers are connected at medium voltage levels. Industrial processes use electrode boilers for producing superheated steam at high pressure.

Electric resistance heating systems convert electric current directly into heat. The heating units may have internal thermostats or external control systems and sometimes even use smart technologies (e.g., programmable scheduled heating) to regulate the temperature. Electric resistance heaters start quicker than heat pumps and require less space and investment costs. However, heat pumps have higher efficiency.

In this study, hybrid heating systems refer to heat pumps coupled with an electric boiler or electric resistance heater. The output of heat pumps depends highly on the outside temperature. When the outside temperature is low, the heat pump performs with lower efficiency as it needs to extract the same amount of heat from cooler air. An electric boiler or resistance heater in tandem with a heat pump helps reduce the problem. Hybrid heating systems can be a promising alternative in terms of flexibility (e.g., heat pump and electric boiler) but require higher installation costs than the individual technologies.

CHP plants consume fuel to produce both power and heat. A recent cross-country analysis using data from 35 countries showed that the CHP share in total electricity generation increases with rising VRE shares [9]. Therefore, CHP is a prominent candidate for bridging between power and heat sectors in the energy transition process. In addition to increased energy efficiency, CHPs offer cost-saving in operation, reduced air pollution, high reliability, improved power quality, and higher productivity. However, it may become difficult for the CHP plants to get enough full load hours when the share of VRE grows. Thus, they will benefit if they are profitable to run also at reduced full load hours and if they can start up and shut down flexibly, operate at low loads, and change the ratio of the heat and power outputs [10].

The technology readiness level (TRL) of most P2H devices indicates that it is feasible to electrify the heating sector rapidly. However, with the increase of P2H devices, there will be an additional burden on the infrastructure. Most distribution grids are not sufficiently strong to incorporate a surge in both P2H devices and electric vehicles. Similarly, wirings and switchgear in buildings may not be sufficient. Thermal Energy Storage (TES) can play a significant role in achieving future decarbonization goals in Europe, especially in a highly renewable energy integrated system. P2H, coupled with TES, can be a promising option for integrating renewable energy, improving operational efficiency, and providing demand-side flexibility and sector coupling. TES can store energy to be used later for heating, cooling, or electricity generation. Large TES with a district heating network can store more heat and supply heat for long periods, offering better flexibility in the sector-coupled system. TES can help to tackle the following three main challenges [11]:

1. VRE and varying demand patterns of P2H technologies cause additional strain on the electricity grid. TES can mitigate this challenge by storing heat;
2. Solar-based heaters generate heat only during the daytime and mostly in summer. Short-term TES can provide stored heat during nighttime, and long-term TES can provide heat during winter, which can help reduce this time-constraint problem; and
3. TES can store thermal energy on large scales to help address daily and seasonal variability in supply and demand for electricity, heating, and cooling. It can help balance the mismatches between CHP operations and the needs of the electricity sector.

## Research scope

Implementing P2H in energy systems will benefit from technical advancements (both hardware and software), changes in policies and regulations, and proactive roles by the involved stakeholders. Improvements in hardware include improving the different system components, upgrading and enhancing network infrastructures (electrical and district heating), and better control and metering systems. On the other hand, software advancements include designing, developing, and enhancing optimization, aggregation, and real-time communication energy system models [8].

In addition to heat pumps, electric boilers, and hot water tanks, other P2H and TES technologies will play an essential role in decarbonizing future energy systems. The mathematical formulation for some of these P2H and TES technologies are available in different energy models. These formulations vary based on various objectives, such as cost minimization, welfare maximization, residual load variability minimization, or flexibility maximization. Most of the models follow cost minimization objectives, assume perfect competition, and use linear programming (LP) or mixed-integer linear programming (MILP) to carry out the optimization [12]. Some of these models exhibit explicit mathematical equations representing different P2H and TES technologies. These modeling methodologies will be examined in the latter part of this paper.

Several recent studies review the potential of P2H. Schweiger et al. studied the possibility of P2H in Swedish district heating systems [6]. They estimated the P2H potential of Sweden to be 0.2 – 8.6 TWh. Similarly, Böttger et al. studied the P2H potential of German district heating grids [13]. According to their estimation, the maximum theoretical P2H potential of the German district heating grid is 32 $GW_{el}$. Hers et al. presented the prospect of P2H in district heating, industry, and horticulture in the Netherlands [7]. They concluded that the techno-economic potential for P2H application in the mentioned sectors could be as high as 3.1 GW. Yilmaz et al. analyzed the future economic potential of flexible P2H in Europe [14]. In another study, Yilmaz et al. analyzed how P2H can increase the flexibility of the European electricity system until 2030 [15]. Bloess et al. reviewed the residential P2H technologies and presented their model-based analyses and flexibility potentials [12]. They pointed out that P2H technologies could cost-effectively contribute to replacing fossil fuel, integrating renewable energy, and decarbonizing the energy system. Ehrlich analyzed the decentralized P2H as a flexible option for the German electricity system [16]. Leitner et al. presented a method for the technical assessment of P2H to couple the electricity distribution systems with local district heating [17]. Kirkerud et al. analyzed how the use of P2H in the district heating system impacts the VRE resources of the Northern European power system [18]. Their results showed a significant increase in VRE market value with an increased installed P2H capacity. Kuprat et al. presented the role of P2H as a flexible load in the German electricity network [19]. Gjorgievski et al. gave an empirical review of the P2H demand response potential of 34 large-scale projects worldwide [20]. Besides P2H, Sarbu et al. presented a comprehensive review of TES [21]. They described the principles of various energy storage techniques and the analysis of storage capacities. Pfleger et al. gave an overview of TES in their study [22], while Enescu et al. reviewed the emerging trends of TES for grid applications [23]. Enescu et al. also addressed the TES models, their characteristics, parameters, and deployment in VRE-based energy systems.

The literature review shows that the potential of P2H is hardly discussed from the perspective of interconnected future European energy systems with a high share of VRE; instead, it is often addressed from a national point of view. The relationship between P2H and TES for providing flexibility as dispatchable loads need further attention in energy research. The literature review also shows that only a few studies characterized the P2H technologies. Bloess et al. classified the residential P2H options [12]. Pieper [24] presented a general overview of household and industrial P2H based on Beck and Wenzl [25]. Schüwer and Schneider presented a similar classification focusing on the industrial sector [26]. Den Ouden et al. characterized P2H for process industries [27]. However, none of these studies explicitly illustrated P2H for all end-use energy sectors.

The most prominent and recent research on model-based analyses of P2H and their modeling formulations is presented in the study by Bloess et al. [12]. They provided a rich set of analytical approaches to implement P2H technologies in power systems and market models. However, their review focused on only residential P2H options. We recognize that the scope needs to be broadened by including the P2H and TES technologies across all sectors and modeling them in energy systems to provide further insights on alternative or complementary decarbonization and flexibility potentials. We also recognize that it is necessary to characterize these technologies and understand their potential roles before presenting these technologies' general mathematical formulations. We limit the presentation of modeling formulations to optimization-based energy models because of their quick and efficient objective seeking within highly complex systems, and their ability to capture sectoral interactions leading to cross-cutting insights [28].

Based on the research scope, the main contributions of this paper are as follows:

1. It identifies and classifies the main P2H and TES technologies, looking across all sectors of the future carbon–neutral European energy system;
2. It briefly describes the technologies and addresses the potential roles in the European context; and
3. It presents the optimization energy modeling equations of these technologies suitable for large-scale energy models.

For the literature review, we screened scholarly databases for articles mentioning relevant keywords of the following topics: power-to-heat, thermal energy storage, combined heat and power, and modeling of these three components. The screening includes journal articles from Energy, Energies, Applied Energy, Energy Policy, International Journal of Energy Research, Applied Thermal Engineering, Renewable and Sustainable Energy Reviews, Sustainability, International Journal of Sustainable Energy Planning and Management, and Energy Economics. Additionally, we reviewed several important articles and reports from different projects in the field of energy systems. The modeling equations are mainly presented for large-scale energy models, which can be modified to suit smaller energy systems. The study does not include a detailed analysis of all available TES technologies; instead, it focuses on thermal storage technologies coupled with P2H technologies. Since CHP may have a considerable impact on linking power and heat sectors, it is considered part of the study. Nevertheless, the paper does not claim to deliver a complete account of all published research on P2H, TES, and CHP. Instead, the study aims to present a comprehensive understanding of significant findings and possible approaches to modeling P2H and TES components in a highly renewable European energy system.

The rest of the paper is structured as follows. Section 2 presents classifications of P2H and TES technologies. Section 3 provides a short description of the main identified P2H and TES technologies and discusses their potential role in the context of the European energy transition. Section 4 provides the modeling formulation for the key P2H and TES technologies for large-scale optimization energy models. Finally, section 5 concludes with some remarks. Furthermore, we have added a Supplementary Material to the paper, which presents the relevant studies, merits, and demerits of the P2H and TES technologies.

## Classification of power-to-heat and thermal energy storage

This section has classified P2H and TES technologies based on existing studies. First, the P2H technologies are categorized based on sectors and temperature levels. Next, TES classification is presented, where thirteen promising technologies are categorized under four major classes.

### *Power-to-heat classification*

Bloess et al. proposed a new classification of residential power-to-heat [12]. They divided the built environment technologies into centralized and decentralized options. The centralized options use district heating, and the decentralized options use individual or community-based local networks. Their classification also indicates storage provisions. For example, centralized P2H comes with storage options, while decentralized P2H may or may not come with storage. Storage can be internal or external hot water tanks. Apart from active thermal energy storage, there can also be passive thermal storage where building mass or interiors store energy. Pieper [24] described an overview of P2H technologies based on Beck and Wenzl [25], where the author identified thermal energy storage as an integral part of P2H to supplement and simplify the operations. Schüwer and Schneider classified the P2H options based on household, trade, commerce, and service (TCS), and industrial heat applications [26]. Household and TCS applications included resistance heating systems, electrode boilers, electric heat pumps, and hybrid heating systems. Industrial P2H had conductive resistance heating, inductive heating, high-frequency heating, magnetic direct current heating, electrical infrared heating, electrode boiler (with and without CHP), and electric heat pumps. Den Ouden et al. presented the electrification strategies and promising technologies for the Dutch process industries [27]. According to their research, while some technologies are already commercially available, other promising technologies are currently in the research phase. Still, they are likely to play essential roles in industrial P2H options.

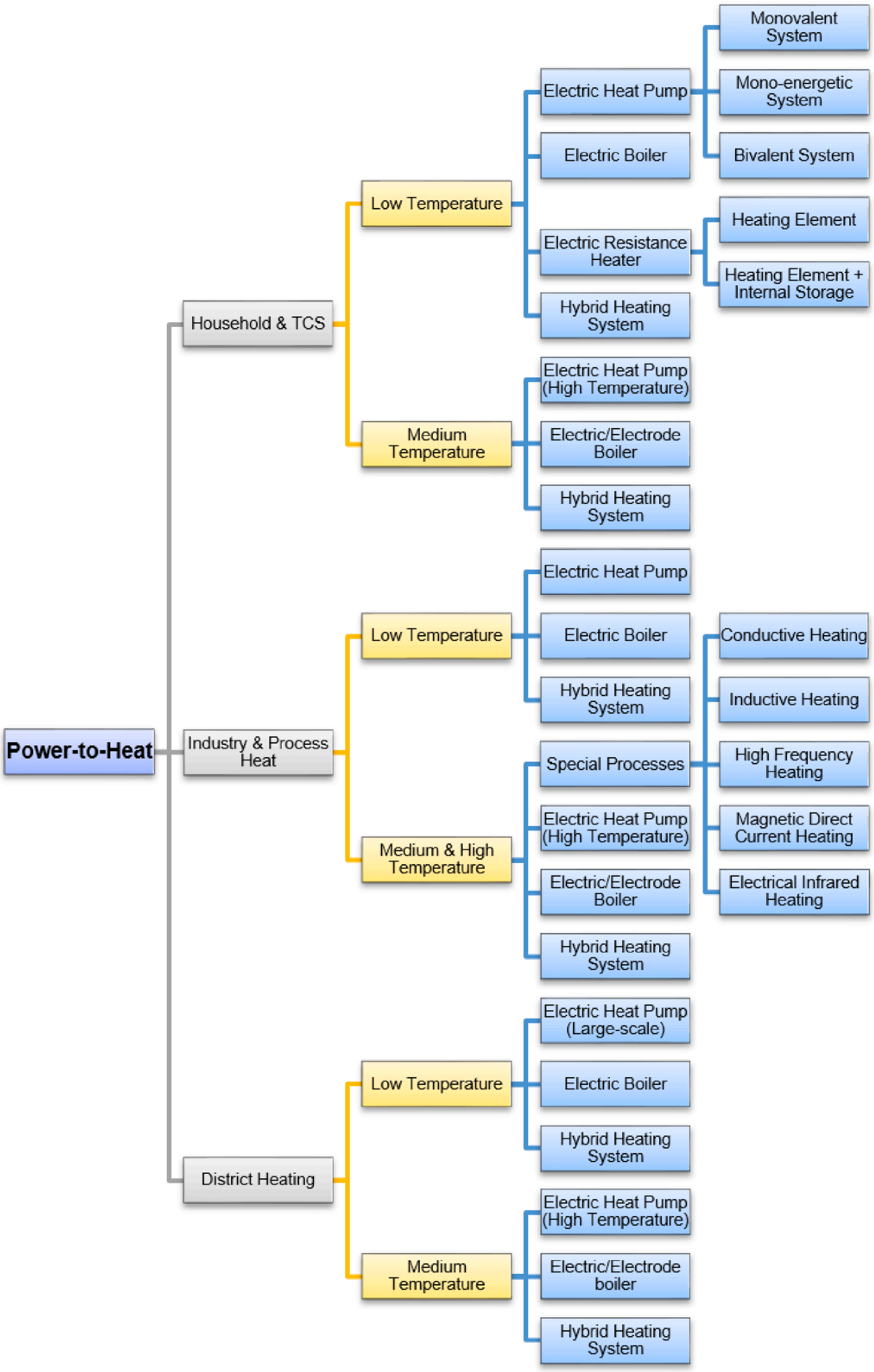

**Fig. 2.** P2H classification.

Fig. 2 illustrates the state-of-the-art classification of different P2H technologies based on the review. First, the P2H technologies are categorized by three sectors: households and TCS, industry and process heat, and district heating. The second classification level is based on temperatures. Garcia et al. presented three temperature levels: (i) low (<120 °C), (ii) medium (120–1000 °C), and (ii) high (>1000 °C) [29]. However, this classification assumes 100–1000 °C as the range for medium temperature. On the third classification level, the technologies are characterized as available technologies. Electric heat pumps and electric boilers are most common. Electric resistance heaters are available mainly in household applications. The medium-temperature applications require high-temperature heat pumps. Large-scale implementations of electric heat pumps are available in district heating applications. Electrode boilers are used in medium and high-temperature applications, especially in high-temperature industrial process heating applications. Hybrid heating systems by combining different P2H technologies can be used in all sectors for different temperature levels. Industries sometimes require special process heaters for high-temperature applications, shown as subtypes in Fig. 2.

Heat pumps appear as one of the most promising P2H technologies[1] [12]. High-temperature and large-scale heat pumps are also becoming popular in industrial process heating and district heating applications. Electric heat pumps can be distinguished as monovalent, mono-energetic, and bivalent options. The monovalent systems have only heat pumps, the mono-energetic systems consist of heat pumps and heating elements, and the bivalent systems consist of heat pumps and auxiliary boilers [12]. Several studies review different heat pump configurations and their design, operation, recent developments, and application potentials to characterize various aspects of heat pump technology. For example, Staffell et al. reviewed the residential heat pumps, focusing primarily on air and ground heat pumps from the UK and Germany [30]. Chua et al. reviewed the recent developments in heat pump systems and analyzed their suitability for various applications [31]. Fischer et al. presented model-based flexibilities of domestic heat pumps [32]. Arpagaus et al. presented a study on high-temperature heat pumps, where they reviewed the market and application potentials in detail [33]. The vapor compression electric heat pump is the most widely used technology among different heat pump configurations, especially for low-temperature applications, because of its simple structure and low initial cost. There are four main components in a vapor compression heat pump: evaporator, compressor, heat exchanger (condenser), and an expansion device. First, the liquid working fluid (i.e., the refrigerant) gets evaporated in the evaporator at low pressure using the heat source (e.g., air, ground, water). Then the vapor is compressed in the electricity-driven compressor, increasing the temperature and pressure of the steam. After that, the high temperature and high-pressure steam enter the heat exchanger, where the heat transfers to the sink. Finally, the condensed vapor goes through the expansion device, where it returns to liquid form, and the cycle repeats. A commonly used indicator for measuring heat pumps' performance is the Coefficient of Performance (COP), calculated from the ratio of heat output and electrical input. COP represents the steady-state performance under a set of controlled conditions with defined input and output temperatures [30].

The second promising option for P2H is an electric boiler. It can be used as simple direct heating resistance boilers in low-temperature cases and complex three-phase electrode boilers in medium and high-temperature cases. Both cases have been widely analyzed[1] [12]. Electrical resistance boilers use an electric heating element that acts as resistance. Electrode boilers use the conductive and resistive properties of water. Other than these two, there are also small-scale infrared and induction-type electric boilers.

Electric resistance heating is another promising P2H option[1] [12]. Electric resistance heating systems use heating elements to generate heat using the Joule effect, where the energy of an electric current is converted into heat as it flows through a resistance. As opposed to electric boilers, no hot water or steam is used. These P2H heaters are used in households and industrial heating systems using conductive, inductive, high frequency, and infrared processes.

Hybrid heating systems generally refer to combining a heat pump with an electric boiler or an electric resistance heater. This combination is typical for low-temperature hybrid heaters in households. However, a different combination of other P2H technologies is also possible. For example, electrode boilers are often combined with CHPs to form hybrid heating systems in district heating systems. Supplementing the heat pump with a gas boiler is another plausible solution. In this way, we can reduce the cost (compared to CHP) and improve the flexibility (compared to direct electric heaters). Furthermore, we can replace gas with electro fuels to avoid emissions in the future.

In addition to the P2H technologies mentioned above, we identified five different industrial process heating systems: conductive heating, inductive heating, high-frequency heating, magnetic direct current heating, and electrical infrared heating [24–27]. Detailed descriptions of other commercial and research-phase process heating systems can be found in [27].

CHP is a mature and proven technology that plays a vital role in integrating power and heat sectors and is likely to be widely used in medium to high-temperature cases in future energy systems[1] [12]. Generally, the two main types of steam turbines in CHP are the non-condensing or backpressure turbine and the condensing or extraction turbine. The backpressure turbine CHP produces electricity and heat with a fixed ratio. The second type is the extraction turbine, where the ratio of electricity and heat can be altered by varying the amount of heat taken from the extraction valve and the amount of energy directed to a low-pressure turbine. Therefore, extraction turbines can offer more flexibility to the system and for the plant operator [34]. Other configurations such as CHP plants with turbine bypass systems are also possible.

## *Thermal energy storage classification*

Thermal Energy Storage is a proven concept used to balance supply and demand for electricity, heating, and cooling. The integration of TES with P2H and CHP applications can provide flexibility and increase the power system's reliability. Most P2H technologies generally combine with external TES. The electric resistance heating systems and some industrial process heating systems that use direct electricity conversion to heat do not need any storage. TES is classified and discussed in most of the literature based on the technologies: sensible heat storage (SHS), latent heat storage (LHS), and thermo-chemical heat storage (THS)[1] [21]. In addition to these three, a study by IRENA identifies thermo-mechanical energy storage (TMS), also known as mechanical-thermal coupled storages, as another promising TES technology [35].

SHS is the most widely used TES form, which stores heat by heating solid or liquid mediums such as water, molten salt, rocks, sand, etc. The term 'sensible' indicates that the medium's heat causes a change in its temperature. Hot water-based SHS is the most used TES because of its low cost, compactness, scalability, maturity, availability, usability, and non-toxicity.

LHS is another popular TES that uses phase change materials (PCM) to absorb and release energy with a physical state change. Promising and commonly used PCMs for TES applications include salt hydrates, fatty acids and esters, and various kinds of paraffin [21]. The term 'latent' indicates that the storage material changes its state (e.g., solid to liquid) for the addition or removal of heat. Using the LHS system with PCM has the advantage of high energy storage density and the isothermal nature of the storage process [21]. Therefore, integrating the heat pump with latent TES can be advantageous, providing constant temperature and compactness to the system.

THS uses thermochemical materials to store and release heat by a reversible endothermic or exothermic reaction process. The most prominent THS materials are water paired with silica gel, magnesium sulfate, lithium bromide, lithium chloride, and sodium chloride. THS can be used to control heat and humidity by using the thermo-chemical adsorption process [21]. This technology is potentially highly efficient (up to 100 %) [11].

TMS is based on mechanical and thermal energy transformations, where the TES is internally combined with mechanical energy storage. TMS includes standard mechanical components such as heat exchangers, compressors, or turbines, sometimes with necessary modifications [36]. The benefits of TMS include the provision to electric and heat storage, high integrability with other heat sources and power generation systems, lower geographic constraints and environmental impacts, and a long lifetime [37].

Based on the IRENA study, we identify thirteen promising TES technologies which can help integrate more VRE into the energy system, as shown in Fig. 3.

Tank Thermal Energy Storage (TTES) uses water as the storage

[1] Please check the supplementary material for all relevant references.

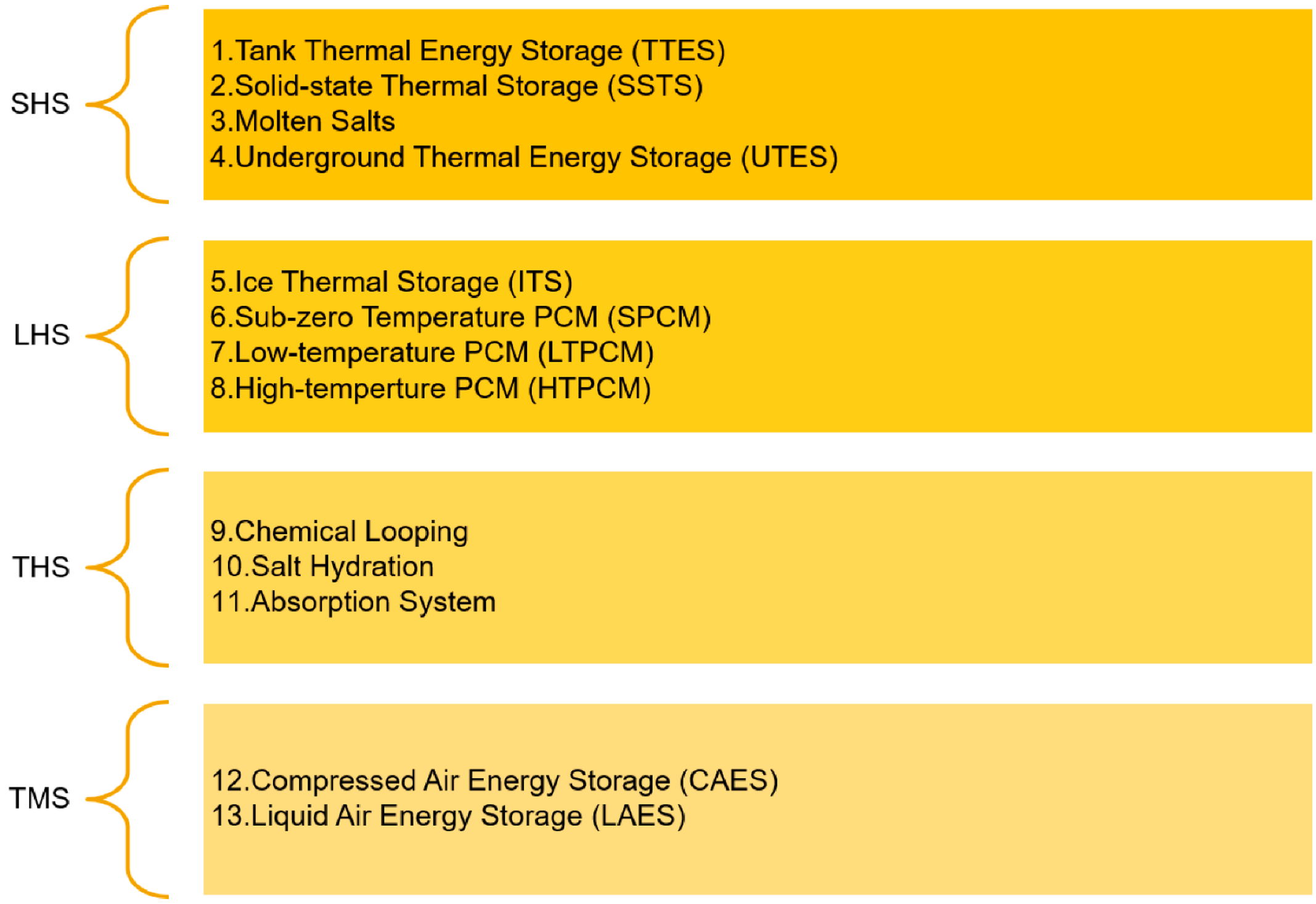

**Fig. 3.** Promising TES for VRE integration. Based on [35].

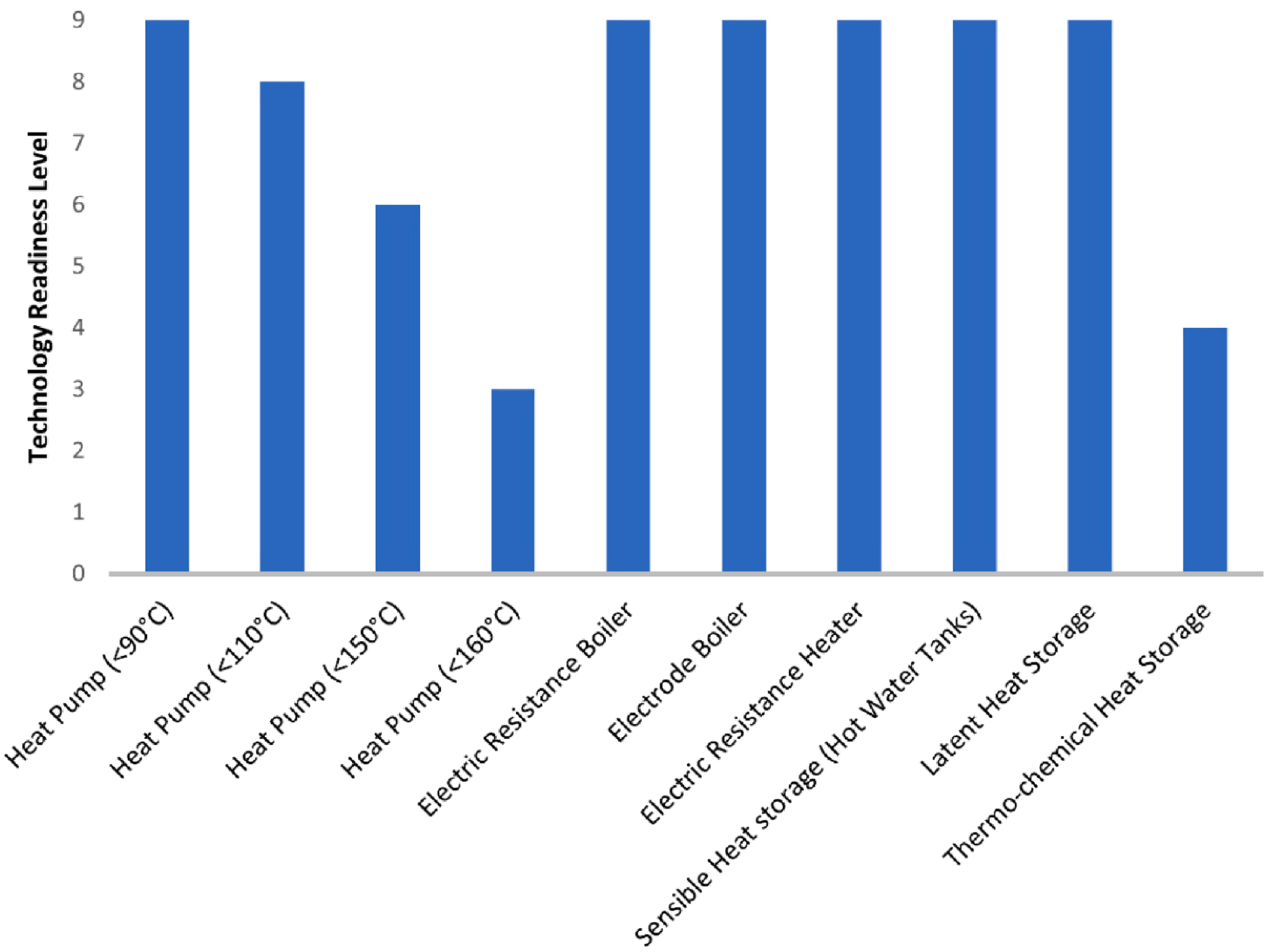

**Fig. 4.** Technology readiness levels of P2H and TES technologies. Data Source: [24].

medium. Solid State Thermal Storage (SSTS) uses ceramic bricks, rocks, concrete, or packed beds as storage medium. Molten salts are inorganic chemical compounds. Underground Thermal Energy Storage (UTES) uses geological strata made up of soil, sand or solid bedrock, or water in artificial pits or aquifers. Ice Thermal Storage (ITS) uses cold energy in ice. Sub-zero Temperature PCMs (SPCM) are single components or are composed of a mixture, such as eutectic mixtures (e.g., salt-water). Low-temperature PCM (LTPCM) uses paraffin waxes and inorganic salt hydrates. High-temperature PCM (HTPCM) uses inorganic salts with high phase-change temperatures. Chemical looping is primarily identified as a potential carbon capture technology using calcium. Salt hydration absorbs and releases energy through hydration and dehydration of solid salts such as magnesium chloride and sodium sulphide. Absorption systems are based on the principle of a concentrated refrigerant solution. In Compressed Air Energy Storage (CAES), the air is stored at high pressure, and in Liquid Air Energy Storage (LAES), it is stored in a liquid form. Adiabatic CAES systems can improve the overall efficiency where an additional high-temperature TES is added.

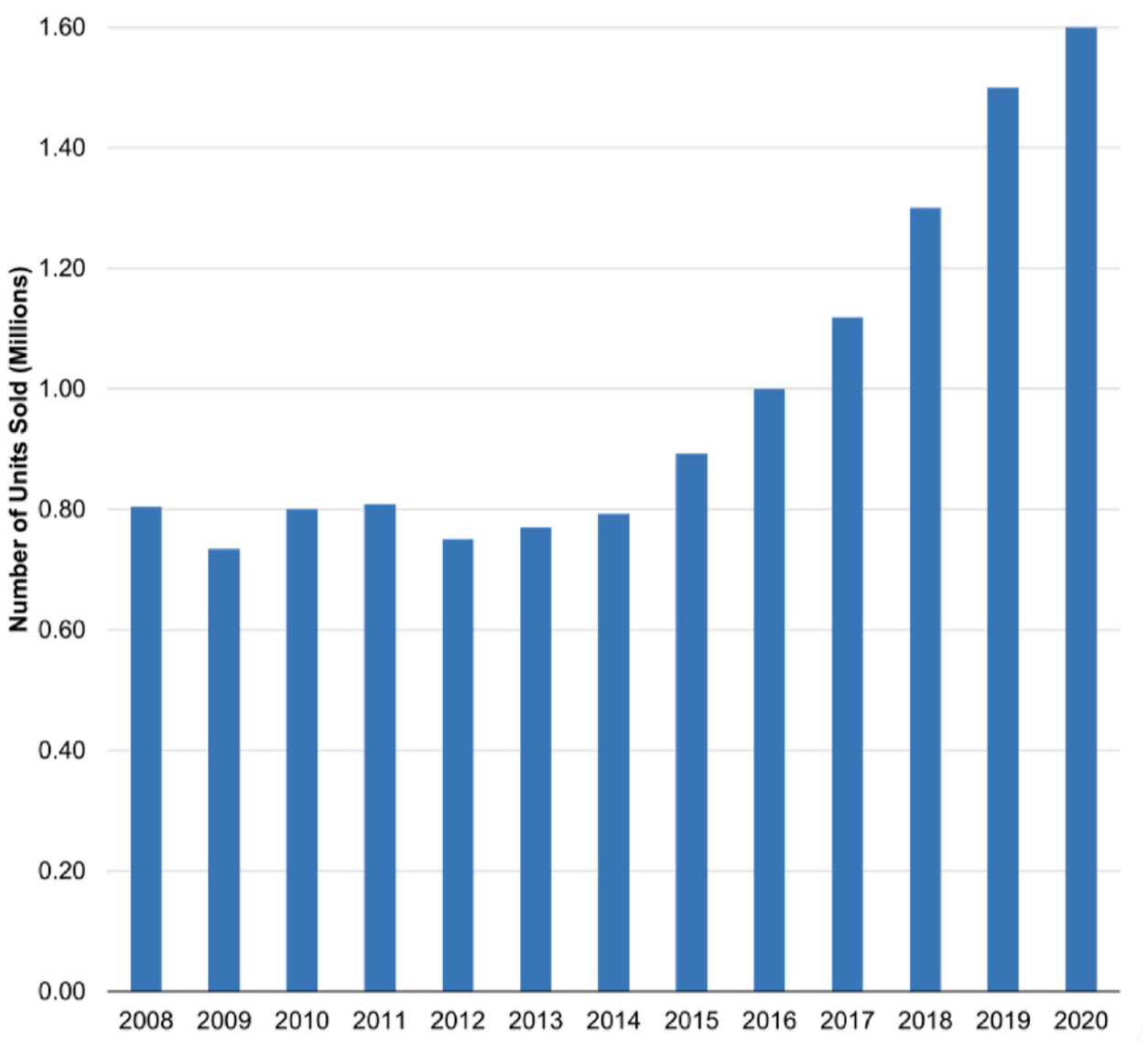

**Fig. 5.** Heat pump sales development in the EU. Source: European Heat Pump Association [39].

Many studies also considered and modeled passive heat storage (PHS) using the buildings' thermal mass, and electrical thermal storage (ETS) using insulated thermal bricks (i.e., electric storage heaters). PHS can store heat in the building's enclosed structure [23] and directly influences the operation of P2H applications (e.g., heat pumps) in the built environment. On the other hand, ETS is mature and advantageous because of its easy management with dynamic charging and comparatively lower cost than batteries [23]. Generally, PHS and ETS store the medium's heat without phase transition and therefore belong to SHS. However, Heier et al. showed that passive TES in buildings can store heat in latent thermal mass [38].

## Potential of power-to-heat and thermal energy storage

Technology readiness levels can help understand the potential of technologies required by the energy transition [24]. Fig. 4 shows the TRLs of prominent P2H and TES technologies based on data from [24]. The ranking scales range from 1 to 9, where level 1 indicates that the technology is in the elemental level of research, and level 9 indicates that the technology is thoroughly tested and proven. Appendix A presents a brief definition of all technology readiness levels. Low-temperature electric heat pumps (<90 °C), electric boilers (resistance and electrode types), and electric resistance heaters are established technologies and are fully technologically ready (TRL 9). For heat pumps over 90 °C, the TRL levels decrease from 9 to 3 as the output temperature increases from <90 °C to <160 °C. Although sensible and latent heat storages have a high TRL (TRL 9), the LHS technology is relatively immature in the domestic segment. On the other hand, thermo-chemical heat storages are still in the early stages of development (TRL 4). In the following sub-sections, we briefly discuss the potential roles of prominent P2H and TES technologies in the EU.

### *Heat pump*

The growing potential of heat pumps is observable in Fig. 5, which shows its sales from 2008 to 2020 in the EU, including the UK (21 countries) [39]. However, the adoption of the different heat pump technologies varies across European countries. The data in Fig. 5 includes all heat pump technologies providing heating, cooling, sanitary hot water, and process heat. The graph shows an upward trend in heat pump sales, achieving more than, on average, 12 % growth per annum in the last six years. According to IRENA, heat pumps will supply 27 % of the total heat demand in the EU by 2050, when the total installation rises to 250 million units in the building sector and 80 million units in the industry sector [8].

High-temperature heat pumps (HTHPs) are becoming increasingly popular in relevance to industrial heating processes. The closed-cycle compression heat pumps are the most widely used HTHPs, while thermally-driven sorption cycles and hybrid absorption-compression heat pumps are the other relevant HTHP technologies. Fig. 6 gives an overview of industrial processes in different industrial sectors identified as suitable for integrating heat pumps. Four color bands indicate four different TRL for the available heat pump technologies. Overall, the food, paper, and chemical industries show promising potential for HTHPs. While drying, pre-heating, boiling, and pasteurization applications can already use commercially available HTHPs (<100 °C), higher temperature heat pumps (>100 °C) are expected to be technologically ready within the next few years [33].

Recent studies showed an immense market potential for industrial heat pumps. Nellissen and Wolf estimated technical potential of 626 PJ, of which 113 PJ is accessible by industrial HTHPs (100–150 °C) [40]. A model-based analysis by Wolf and Blesl showed that the technical potential of heat pumps across the industrial sector in EU-27 and the UK is 1717 PJ [41]. However, only 15 % of the potential (270 PJ) is accessible because of economic considerations. Another study by Marina et al. estimated that industrial heat pumps up to 200 °C can cover 641 PJ of the process heat demand of EU-27 and the UK [42].

### *Electric boiler*

Electric resistance boilers are usually connected at low voltages (e.g., 400 V) and have a low capacity (<5 MW). Electrode boilers are generally connected at medium voltage levels (e.g., 10 kV) and have a higher capacity (3–70 MW). They can produce superheated steam with high temperatures (>350 °C) and high pressure (>70 bar). Both types of electric boilers have high efficiency ranging from 95 to 99.9 % [27]. Table 1 shows the industrial applications of electric boilers according to various temperature levels.

In the residential sector, the electric boilers are often identified as a supplementary option [12]; however, they have a higher potential in district heating networks and industries [43]. We did not find any study that estimated electric boilers' potential using energy models. Nonetheless, the rising trend of electric boiler usage in the past few years can be seen in Eurostat data [44]. As shown in Fig. 7, the electricity consumption by electric boilers for all sectors increased from 243 GWh to 697 GWh in the last ten years (EU-27 and the UK) [44]. The trend indicates plausible higher usage of electric boilers in the future, especially in an energy system with a highly electrified industrial sector.

### *Electric resistance heater*

For households, electric resistance heating using only heating elements can be supplied by radiant heaters, baseboard heaters, wall heaters, underfloor heating systems, and electric furnaces. Heating elements can also be associated with internal ceramic blocks, known as electric storage heaters. An electric storage heater is a flexible P2H application that can reduce the peak demand by storing heat in ceramic blocks at low price times.

In industrial processes, an electric process heater is a form of resistance heating that is technologically matured and can be used in high temperature and pressure applications. These heaters use several heating elements across which the working fluid flows lengthwise and crosswise to be heated up to 600 °C. Electric arc furnaces and the Hall-Héroult process are proven conductive heating processes for metallurgy applications [24].

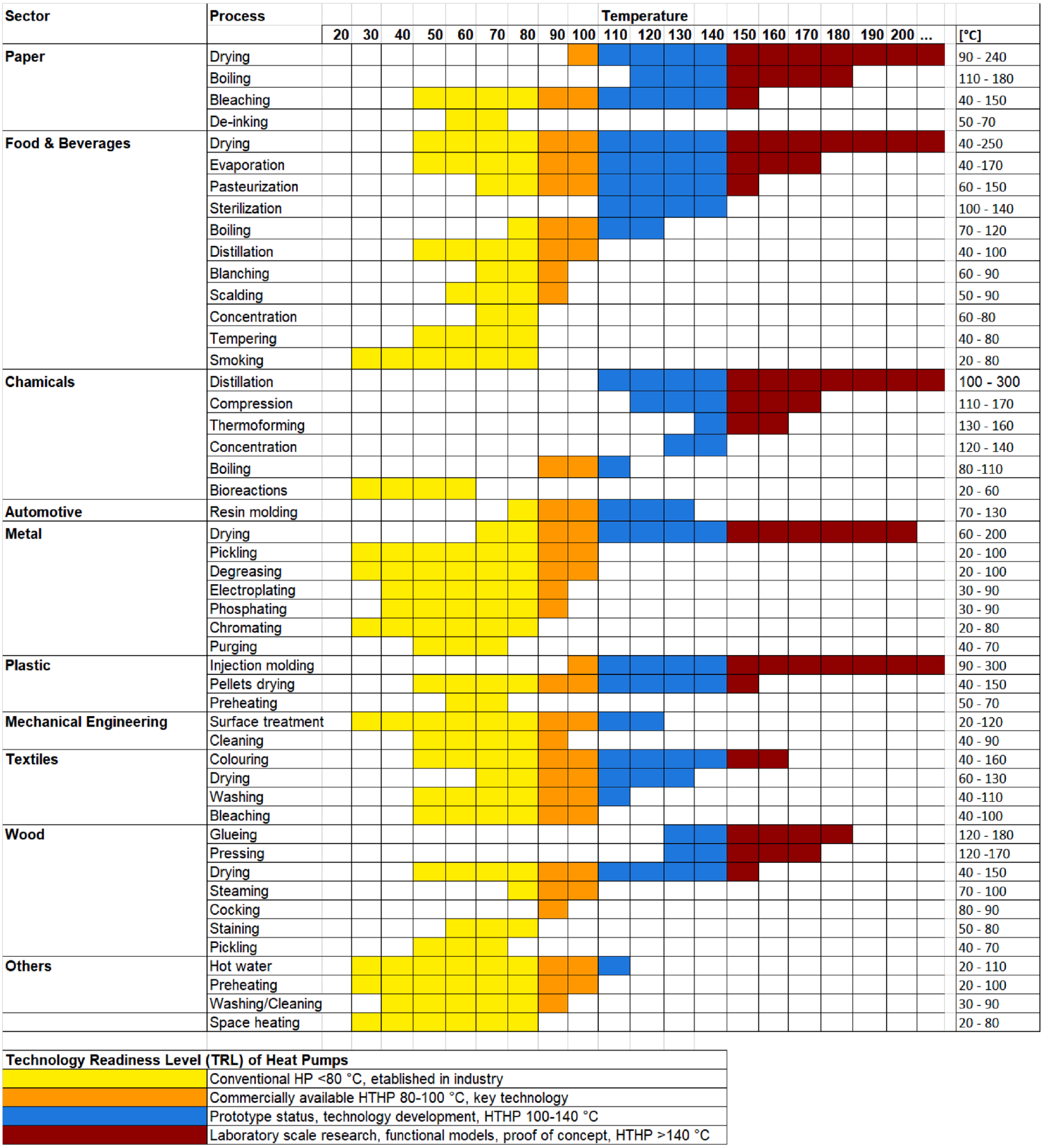

**Fig. 6.** Overview of industrial processes suitable for heat pumps. Source: [33].

**Table 1**
Temperature-wise industrial applications of electric boilers. Source: [15].

| Applications | Temperature Level | Use in industries |
|---|---|---|
| Low-temperature | <120 °C | Food & beverages, chemicals, textiles, dairy, breweries, mineral oil, etc. |
| Medium-temperature | 120–1000 °C | Drying, production of plastic materials, plasterboards, bitumen, asphalt, etc. |
| High-temperature | >1000 °C | Process heating such as the production of iron, steel, bricks, cement, etc. |

Besides conductive heating, inductive, high-frequency, and infrared heating systems are also popular and proven in electrifying industrial processes. Inductive heating uses electromagnetic induction to generate heat and is available for several applications such as furnaces for heating metals, welding, cooking, brazing, sealing, heat treatment, and plastic processing. High-frequency heating, also known as microwave or radiofrequency heating, is also commonly used in textiles, paper, food, plastic and chemical industries. Finally, infrared heating is another commercially available P2H application used in various industrial processes such as drying, curing, welding, and coating.

The use of electric resistance heating depends heavily on the energy sources and the countries' energy policies. For example, in countries with a high share of nuclear power, electric storage heaters can store heat using electricity in times of excess generation. We can expect the same for countries with high VRE shares. On the other hand, direct electrical heating is widespread in countries with high hydropower resources or fewer wintery days. Installing decentralized and relatively cheaper electrical heating systems in such countries is more cost-effective than investing in expensive heat pumps or electric boilers. In the future, the development of smart grids and grid interaction with direct electric heaters may lead to positive technology advancements [15].

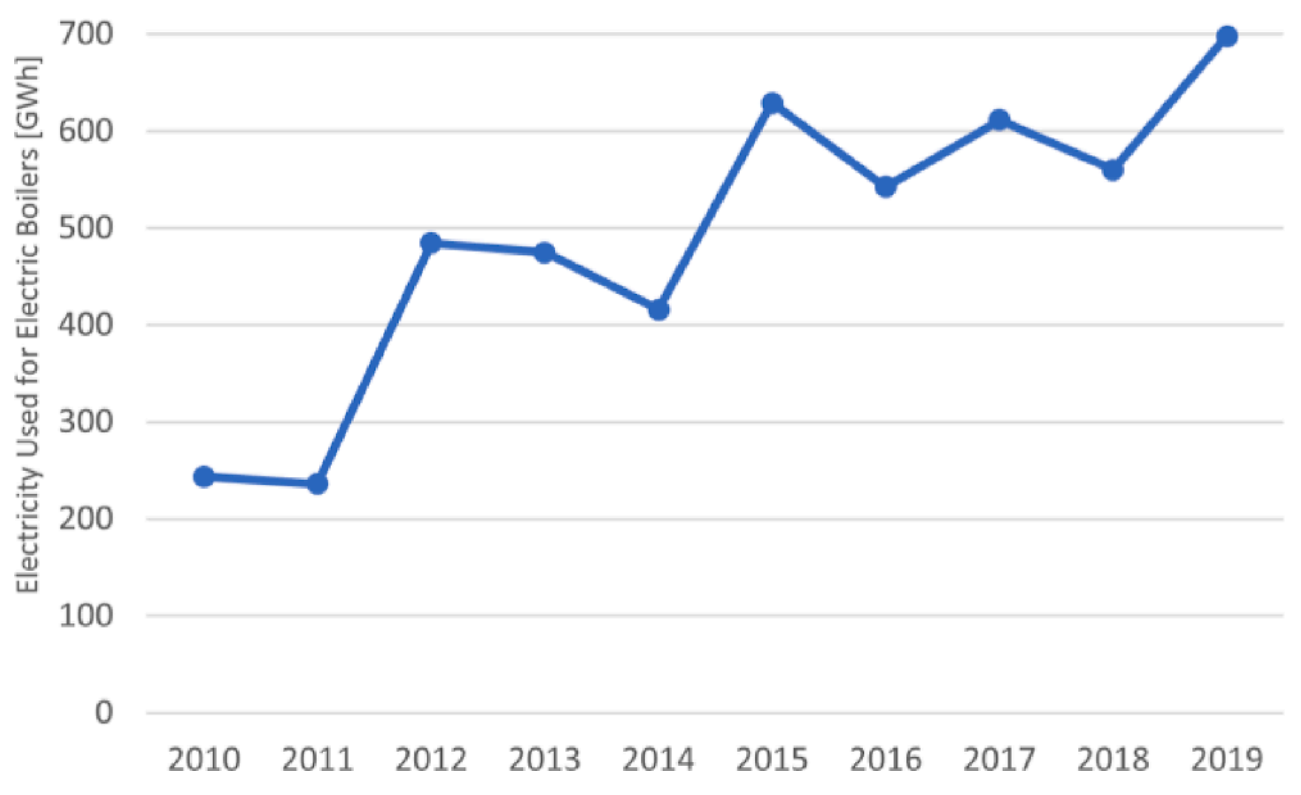

**Fig. 7.** Electricity used by electric boilers in EU-27 and the UK. Data Source: Eurostat [44].

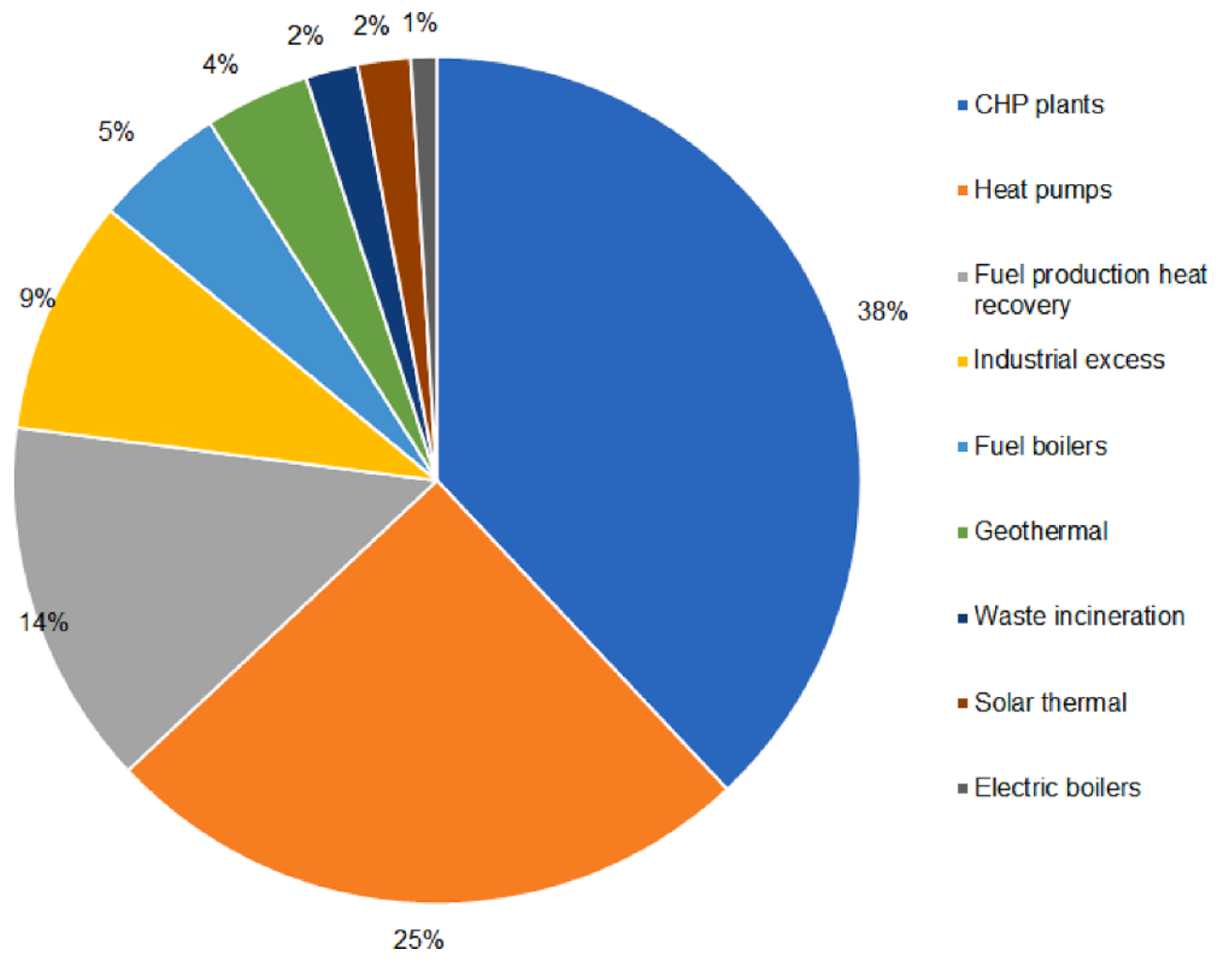

**Fig. 8.** Share of district heating sources according to HRE 2050. Adapted from [49].

*Combined heat and power*

Combined Heat and Power is known as the most efficient technology that can deliver both heat and power simultaneously and can utilize waste and biomass resources [45]. Based on the Eurostat statistics for 2018 in EU-27 and the UK, CHP electricity generation was 327 TWh, and total CHP heat production was 2787 PJ [46]. An optimal combination of CHP with P2H and TES technologies in district heating systems can facilitate flexible sector coupling of power and heat and shows excellent potential to increase renewable shares in the energy system [47,48]. In addition to district heating, CHP can also be used to supply industrial process heating. The JRC policy report in 2017 states that the conversion of existing power plants to CHPs will increase the overall efficiency of the European energy system, which is otherwise limited to 50 % [34]. According to the Heat Roadmap Europe (HRE) 2050 scenarios, district heating should include various sources to ensure flexibility and low emission levels. Fig. 8 shows the distribution of district heating source shares in HRE 2050 for 14 countries in the EU, where CHP covers a significant portion (38 %) of the total district heating [49].

A recent study published by Artelys finds that CHP will play a fundamental role in achieving total decarbonization in Europe by 2050 [50]. According to their analysis, CHPs can maximize system efficiency and flexibility to complement high VRE generation. An optimized CHP deployment can save over €8 billion compared to a lower CHP deployment solution, allowing an annual $CO_2$ emission reduction of up to 5 metric tons. They also found that CHP is suitable in all economic sectors, and it can foster a higher use of biomass resources.

Fuel cell micro-cogeneration is the most contemporary intelligent heat and power solution for residential and small and medium enterprises, potentially saving a considerable amount of energy, reducing CO2 emissions, and lowering energy bills [51]. It is also adaptable and reduces air pollution by efficiently producing energy and heat without combustion. The fuel cell-based micro-CHP technology is ready for widespread adoption from a technical standpoint. Furthermore, according to a life cycle cost analysis, micro-CHP technology can become economically viable. Subsidies can help improve the economics of micro-CHP systems in the short term, and they may be necessary for the technology to reach the mass market.

*Thermal energy storage*

TES is considered an essential tool for smart heating and cooling concepts, playing significant roles in different applications. Residential TES can be used as a demand response tool for energy arbitrage, load variability reduction, and reserve provisions [52]. The flexibility of using residential TES (e.g., smart electric thermal storage) can facilitate the consumers to maximize their local RES usage. The integration of TES in district heating systems can significantly increase system flexibility and facilitate the smooth coupling of P2H technologies in the energy system [6]. TES can improve the overall storage capacity and enhance operational strategies in smart community-based energy systems [23]. Mobile TES offers additional flexibility by making heat available at remote locations [53,54]. Waste heat storage in TES can increase industrial processes' efficiency and operational flexibility. Furthermore, the combination of TES with HTHP can increase the overall energy efficiency, reduce VRE curtailments, reduce system cost, and improve environmental footprint [53–55].

According to a recent study by LUT and EWG, TES will emerge as the most relevant heat storage technology across all sectors in Europe, with around 40–60 % of heat storage output from 2030 until 2050 [56]. Gas storage is expected to cover the rest of the demand. The thirteen promising TES technologies for 2050, identified by IRENA, can be distributed according to their applications across different sectors, as shown in Fig. 9 [35].

## Modeling of power-to-heat and thermal energy storage

Three modeling approaches are prominent in energy system optimization models: Linear Programming, Mixed-integer Linear Programming, and Nonlinear Programming (NLP). Most energy system models use LP because of its inherent computational advantages. In large-scale energy systems with numerous components, LP can serve as a simple, fast, scalable, and straightforward method. Although MILP and NLP allow more accurate modeling results than LP, they can be impractical for large-scale models due to higher computational times. Ommen et al. compared LP, MILP, and NLP in energy models in [57], where they concluded that MILP is the most suitable option considering model runtime and accuracy. Nevertheless, when it comes to large-scale energy systems with detailed characteristics from individual components, LP is a more preferred alternative by most researchers. It is also possible to combine LP results with MILP so that MILP is used for analyzing a minor part of the large-scale energy system in detail [58]. This section presents the mathematical models for the P2H and TES technologies. Through this section, we use capital letters to denote parameters and small letters to represent variables.

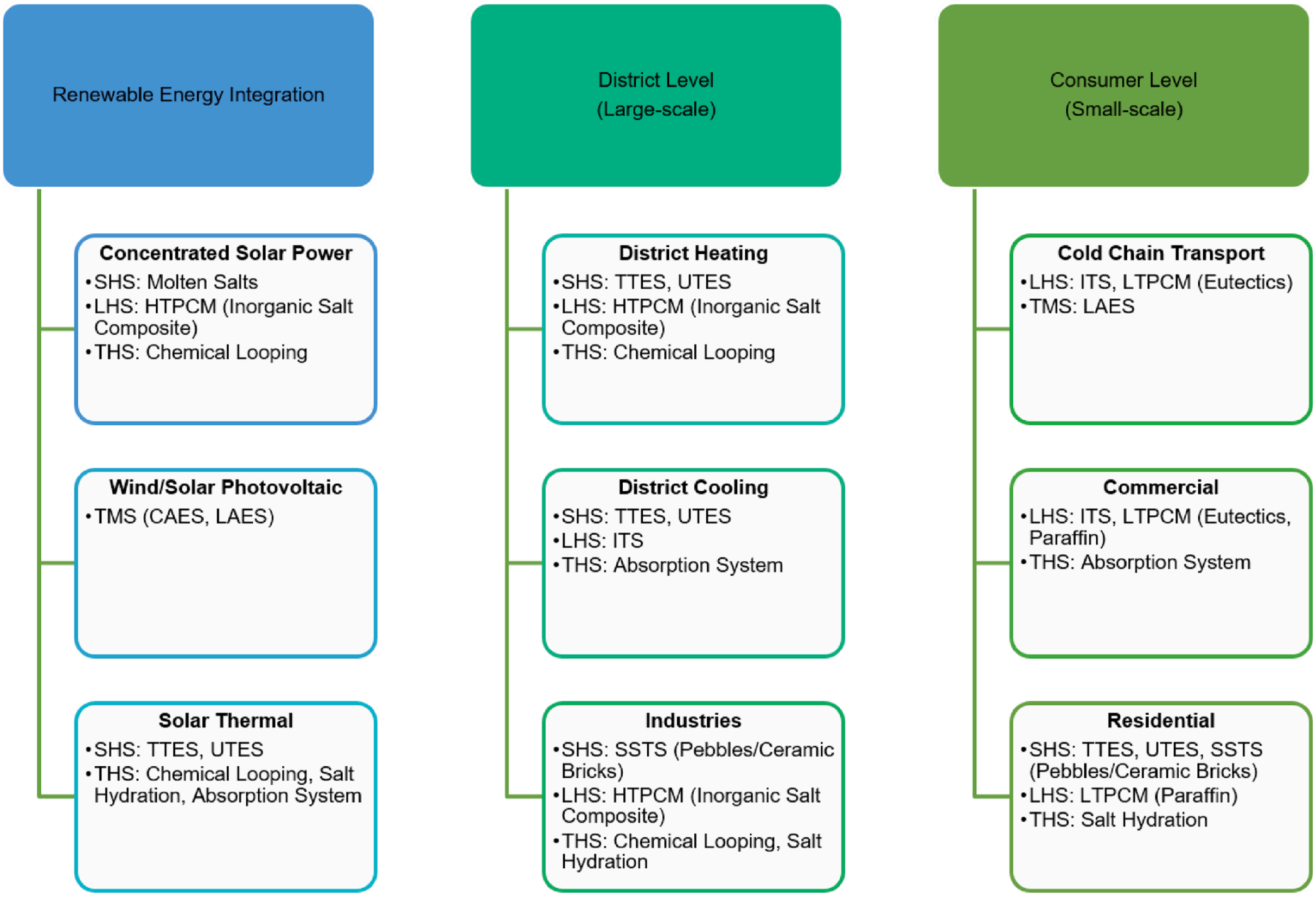

**Fig. 9.** Most promising TES technologies for 2050 and their sectoral distribution according to applications. Based on [35].

*Heat pump*

While modeling heat pumps, it is essential to acknowledge the difference in temperatures of the source and the sink [12,59–62]. The theoretical maximum COP formulates as:

$$COP_{Theory}(t) = \frac{Q_H(t)}{Q_H(t) - Q_L(t)} = \frac{T_H(t)}{T_H(t) - T_L(t)} \ \forall\ t \quad (1)$$

where $Q_H$ denotes the heat supplied to the high temperature reservoir (sink), $Q_L$ denotes the heat supplied from the low temperature reservoir (source), $T_H$ denotes the sink temperature, $T_L$ indicates the source temperature, and $COP_{Theory}$ represents the maximum theoretical COP (i.e., Carnot COP without any loss).

The relationship between the maximum theoretical COP and the actual COP after loss depends upon the efficiency $\eta_{HP}$ of the heat pump. The actual COP is calculated as:

$$COP(t) = \eta_{HP} \bullet COP_{Theory}(t)\ \forall\ t \quad (2)$$

In reality, the sink temperature of a heat pump is directly affected by the heat flow$\dot{q}_{HP}$ . Therefore, the heat pump is modeled as:

$$\dot{q}_{HP}(t) = p_{HP}(t) \bullet COP(t)\ \forall\ t \quad (3)$$

where $p_{HP}$ represents the electrical power input of the heat pump.

The formulation in (3) is the most widely used model in the academic literature on heat pumps in the energy system [12]. The heat pump can operate between a minimum $P_{HP\,_}$ and maximum $\overline{P_{HP}}$ electric power input (4). The heat output is constrained by an upper limit $\overline{Q_{HP}}$ (5). A binary variable $u$ is introduced to model non-linear costs (e.g., no-load cost, start-up cost) and the minimum output.

$$P_{HP\,_}.u(t) \le p_{HP}(t) \le \overline{P_{HP}}.u(t)\ \forall\ t \quad (4)$$

$$\dot{q}_{HP}(t) \le \overline{Q_{HP}}.u(t)\ \forall\ t \quad (5)$$

$$\dot{q}_{HP}(t) \ge 0\ \forall\ t \quad (6)$$

If we consider a variable temperature window for the sink (such as 18 – 22 °C for comfort level), the heat pump can provide flexibility to the system. In that case, the sink temperature $T_H$ becomes a variable, and equation (1), which is embedded in equations (2) and (3), becomes nonlinear. Similarly, $T_L$ often depends on the weather, which can be considered when modeling the COP. However, in large-scale aggregated energy models, it is common to assume a constant COP. Demand response through load shifting can be applied to provide flexibility in the heat pump-based system, which can be modeled using formulations from Morales-España et al. by supplying a given demand in a given maximum delay time window [63].

In the case of reversible heat pumps for air conditioning or refrigeration, the relationship between maximum theoretical COP for cooling $COP^{Cooling}_{Theory}$ and the source and sink temperatures is expressed as [64,65]:

$$COP^{Cooling}_{Theory}(t) = \frac{Q_L(t)}{Q_H(t) - Q_L(t)} = \frac{T_L(t)}{T_H(t) - T_L(t)}\ \forall\ t \quad (7)$$

where $Q_L$ denotes the heat extracted from the low temperature reservoir (i.e., the cooling load), $Q_H - Q_L$ indicates the work required for cooling, $T_L$ represents the sink temperature (low temperature reservoir), and $T_H$ indicates the source temperature (high temperature reservoir).

Other comprehensive heat pump formulations to address the temperature dependence of COP can be found in various studies. For example, Verhelst et al. suggested four different empirical approximations of the physical laws governing heat pump operation based on the data from the heat pump manufacturer [66]. Heinen et al. suggested pre-computing the heat pump dependence on temperature using a linear equation [67]. They determined the slope of the equation from heat pump performance data, assumed a constant ambient temperature of 280.15 K (7 °C) according to EU performance regulations, and fit the relation to a constant COP. Georges et al. presented piecewise linearization of the nonlinear problem from Verhelst et al., requiring empirical manufacturer data and considering nominal conditions [68]. Heat pump formulations by Salpakari et al. involve district heating integration and are limited to a supply temperature of 90 °C [69]. Staffell et al. proposed a generic regression performance map for modeling COP variance, which used surveys of industrial data sheets and field trials [40]. Their formulations apply to household-scale heat pumps as well as large-scale industrial heat pumps [70]. Fischer et al. followed a similar approach in [32], which has been applied to manufacturer data under different temperature conditions by Ruhnau et al. in [71]. To summarize, comprehensive heat pump modeling based on available data is another viable alternative, as it has been used by many researchers. Such methods are preferred in comparatively small-scale and complex heat pump models, such as capturing higher efficiency in varying load modes or allowing higher flexibility in variable speed heat pumps.

*Electric boiler and electric resistance heater*

The relationship between electric power input $p_{EB}$ and heat output $\dot{q}_{EB}$ in an electric boiler is widely modeled as [72,73]:

$$\dot{q}_{EB}(t) = \eta_{EB}.p_{EB}(t)\ \forall\ t \tag{8}$$

where $\eta_{EB}$ indicates the efficiency of the electric boiler. The electric boiler usually operates between a minimum $P_{EB_}$ and maximum $\overline{P_{EB}}$ electric power input (9). The heat output is constrained by an upper limit $\overline{Q_{EB}}$ (10) [74]. The binary variable $u$ is used to model non-linear costs and the minimum output.

$$P_{EB_}.u(t) \leq p_{EB}(t) \leq \overline{P_{EB}}.u(t)\ \forall\ t \tag{9}$$

$$\dot{q}_{EB}(t) \leq \overline{Q_{EB}}.u(t)\ \forall\ t \tag{10}$$

$$\dot{q}_{EB}(t) \geq 0\ \forall\ t \tag{11}$$

Electric boilers and electric resistance heaters are capable of transitioning from no-load to full-load state within minutes or even seconds. Nielsen et al. did not consider any start-up costs and ramping constraints for electric boilers [75]. The modeling of electric boilers can be more complex, taking the thermal stratification effect into account. Thermal stratification in electric boiler storage tanks indicates different temperature levels in several layers inside the tank. In energy system models, many approaches are used to address the thermal stratification effect. Celador et al. used three techniques to model hot water storage tanks: actual stratified, ideal stratified, and fully-mixed [76]. Farooq et al. presented experimental results of a low-pressure domestic electric boiler with eight stratification layers [77]. De Cesaro Oliveski et al. introduced a numerical and empirical analysis of temperature and velocity inside the hot water tank using one-dimensional and two-dimensional models [78]. Diao et al. tested electric boilers' response with various control strategies and considered two modes in a comprehensive model: one node and two nodes [79]. Han et al. presented a review of different types of thermal stratification tanks, their research methods, and compared their efficiencies [80]. In electric boiler modeling, single mass, one-node, or fully mixed tank models are widely used because of their simple formulation. These models consider that the tank's water is mixed and has a uniform temperature without any thermal stratification. This paper considered the single mass system without any thermal stratification for modeling in energy systems. The readers are suggested to go through the references [81–84] for a detailed formulation of stratified boilers.

Electric resistance heaters can be modelled using the same equations (8)–(11), where the ratio between heat production and electricity has different values.

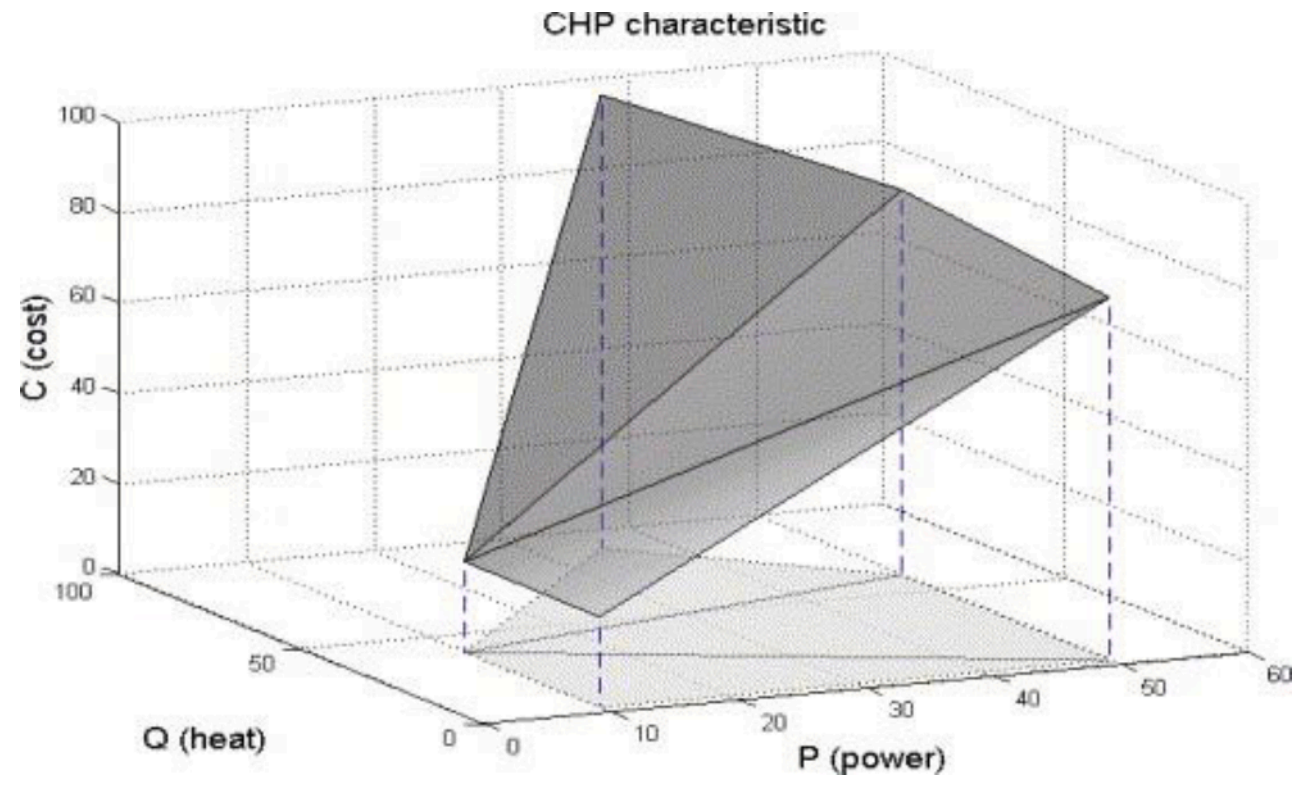

**Fig. 10.** Feasible Operating Region (FOR) of a convex CHP plant. Source: [85].

*Hybrid heating system*

In the case of hybrid heating systems, each technology is modeled individually. For example, if a supplementary equipment complements the heat pump, the relation between the total power required by the hybrid system $p_{HB}$ is the sum of electric power required by the heat pump $p_{HP}$ and the electric power required by the supplementary system $p_{SUP}$:

$$p_{HB}(t) = p_{HP}(t) + p_{SUP}(t)\ \forall\ t \tag{12}$$

Equations (1)–(5) are used to model the heat pump, and (7)–(11) are used to model the supplementary equipment, which is either an electric boiler or an electric resistance heater. Patteeuw et al. modeled hybrid heating systems using such approach [60]. The heat from P2H technologies is part of the total heat demand $\dot{Q}$ of the system (which can be an exogenous parameter), as shown in (13):

$$\dot{q}_{HP}(t) + \dot{q}_{SUP}(t) + \dot{q}_{GB}(t) \geqslant \dot{Q}(t) \forall t \tag{13}$$

where $\dot{q}_{GB}$ denotes the heat from a gas boiler. For the sake of completeness, we assume that there is a gas boiler in combination with the heat pump and the supplementary system. Flexibility to such hybrid systems can be provided either by shifting the sources (e.g., from electricity to gas) or by shifting in time. Time shifting can be enabled by TES, which allows stored energy to be used later, or by a specific technology, such as heat pumps, which can provide shifting in time through demand response formulations, as presented in [63].

*Combined heat and power*

In CHPs, power, heat, and cost depend on each other resulting in a convex feasible operating region (FOR) as shown in Fig. 10 [85].

The operation of a single CHP unit as a convex combination of the extreme points $C_j$, $P_j$, $Q_j$ of the characteristic surface is expressed using (14)–(18) [45,85,86]. $C_j, P_j, Q_j$ indicate the production cost, power production and heat production at characteristic point $j \in J$, respectively. Characteristic points are the extreme points of the operating region of the plant. Variable $x_j$ is used to encode the operating region as a convex combination of extreme points (14)-(16). Variable $u$ indicates the commitment status of the unit for hour $t$, which is equal to 1 if the unit is online and 0 if offline. Parameter $p_{CHP}$ indicates the net power production, $q_{CHP}$ indicates net heat production, and $c_{CHP}$ indicates the production cost of the convex plant. The production cost mainly indicates the fuel cost, however, other variable costs (e.g., maintenance costs) can be included.

$$c_{CHP}(t) = \sum_{j \in J} C_j x_j(t)\ \forall\ t \tag{14}$$

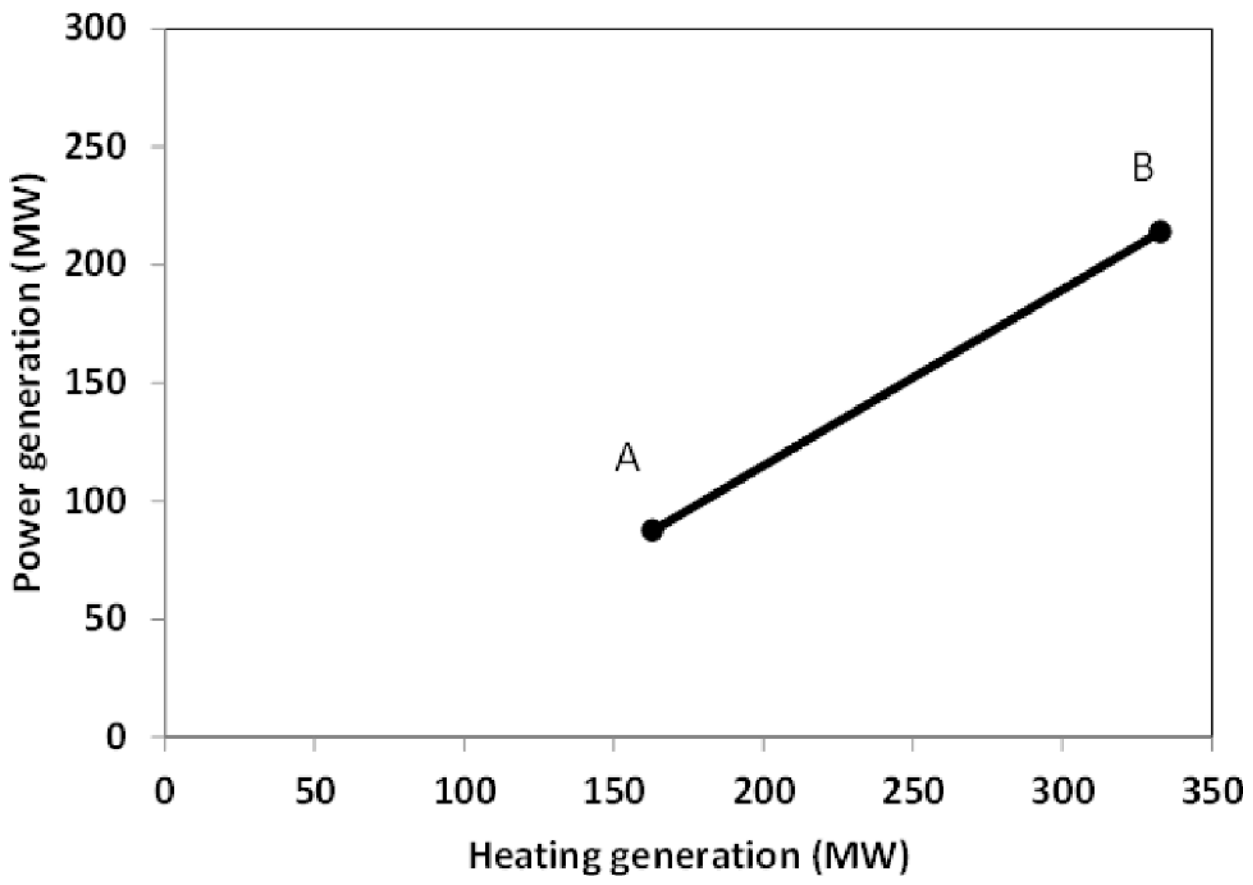

**Fig. 11.** Feasible operation region (FOR) of CHP with a backpressure turbine. Source: [34,74].

$$p_{CHP}(t) = \sum_{j\in J} P_j x_j(t)\ \forall\ t \tag{15}$$

$$q_{CHP}(t) = \sum_{j\in J} Q_j x_j(t)\ \forall\ t \tag{16}$$

$$\sum_{j\in J} x_j(t) = u(t)\ \forall\ t \tag{17}$$

$$x_j(t) \geq 0, j \in J\ \forall\ t \tag{18}$$

The following constraints ensure that the unit operates within the ramp rate limits [87]:

$$p_{CHP}(t) - p_{CHP}(t-1) \leq RU\ \forall\ t \tag{19}$$

$$-p_{CHP}(t) + p_{CHP}(t-1) \leq RD\ \forall\ t \tag{20}$$

Where $RU$ is the ramp-up rate, and $RD$ is the ramp-down rate of the unit. The model can be easily expanded to include spinning reserves [87]. The start-up cost of CHP $c^{SU}_{CHP,g}$ is constrained by the online status of the unit in two consecutive time steps and the start-up cost parameter $C^{SU}_{CHP}$ as shown in (21):

$$c^{SU}_{CHP}(t) \geq (u(t) - u(t-1)).C^{SU}_{CHP}\ \forall\ t \tag{21}$$

Morales-España et al. provide other constraints, such as minimum uptime and downtime, startup and shutdown capability, variable startup cost, etc., that can apply to this model [87]. CHP units can also be modeled using multiple conversion components, particularly when the plant has more operating modes and offers more flexibility [55,88]. Helistö et al. present a more generic formulation where the unit can have multiple inputs and outputs [89]. It also allows the chaining of units to present more complicated plants. Nevertheless, the convex FOR is usually simplified using fewer parameters, as discussed in the following subsections that present formulations for backpressure and extraction turbines.

*Backpressure turbine*

The FOR of a backpressure turbine CHP plant is presented in Fig. 11 [34,74]. The FOR is expressed as (22)–(23), where σ is the fixed power-to-heat ratio, $\overline{P_{CHP}}$ is the maximum power generation, and $\underline{P_{CHP}}$ is the minimum power generation of the backpressure unit [34,74].

$$p_{CHP}(t) = \sigma.q_{CHP}(t)\ \forall\ t \tag{22}$$

$$\underline{P_{CHP}}.u(t) \leq p_{CHP}(t) \leq \overline{P_{CHP}}.u(t)\ \forall\ t \tag{23}$$

Limiting heat generation is not required in CHP mode because it is achieved by the fixed power-to-heat ratio (22) together with (23). In the case of large-scale energy models, $u$ can be relaxed (i.e., a continuous variable between 0 and 1) resulting in an LP approximation of the problem [90].

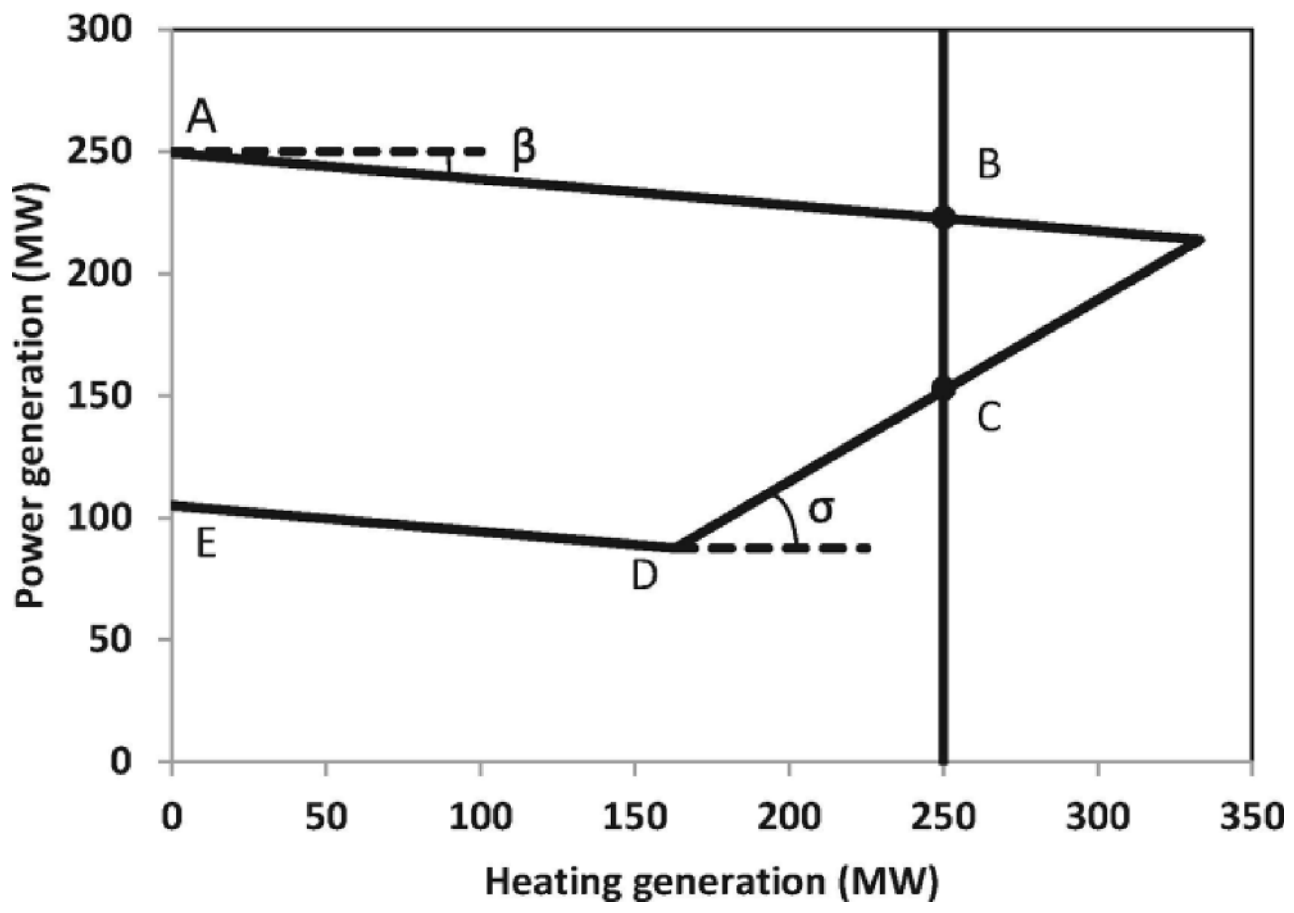

**Fig. 12.** FOR of CHP with an extraction turbine. Source: [34,74].

*Extraction turbine*

The FOR of an extraction turbine CHP plant is presented in Fig. 12, where A-B and E-D indicate the power-loss limits (isofuel lines at maximum and minimum production), D-C indicates the maximum heat limit for a given amount of power, and B-C indicates the maximum possible heat extraction [34,74].

Therefore, the FOR of the extraction turbine is modelled as [34,74]:

$$p_{CHP}(t) \geq \sigma.q_{CHP}(t)\ \forall\ t \tag{24}$$

$$q_{CHP}(t) \leq \overline{Q_{CHP}}.u(t)\ \forall\ t \tag{25}$$

$$p_{CHP}(t) \leq \overline{P_{CHP}}.u(t) - \beta.q_{CHP}(t)\ \forall\ t \tag{26}$$

$$p_{CHP}(t) \geq \underline{P_{CHP}}.u(t) - \beta.q_{CHP}(t)\ \forall\ t \tag{27}$$

where $\overline{Q_{CHP}}$ is the maximum heat generation and $\beta$ is the power loss index (ratio between lost power generation and increased heating generation). Following the FOR of Fig. 12, (24) imposes the limit indicated with the points D-C, (25) imposes the limit B-C, (26) imposes the limit A-B, (27) imposes the limit E-D, and the limit A-E is imposed by the variables being defined non-negative. A collection of typical parameter values from various literature references for this model can be found in [34].

*Thermal energy storage*

Here we present the generic formulations for TES, notably for hot water tanks (sensible heat storage). The tank is assumed to be perfectly stirred; therefore, it has the same temperature range in every layer. In addition, the storage losses are generally taken into account either as only stationary losses [16,61,67,75,91,92], or stationary and dynamic losses [93]. As a result, TES balance is expressed using (28)–(31):

$$s(t) = (1-L_s).s(t-1) + (1-L_D).\dot{q}_C(t) - (1-L_D).\dot{q}_D(t)\forall t \tag{28}$$

$$s(t) \leq S_{cap}\ \forall\ t \tag{29}$$

$$\dot{q}_C(t) \leq \overline{Q_{FLOW}}\ \forall\ t \tag{30}$$

$$\dot{q}_D(t) \leq \overline{Q_{FLOW}}\ \forall\ t \tag{31}$$

where $s$ is the heat storage level, $L_s$ is the stationary heat loss, $L_D$ is the dynamic heat loss, $\dot{q}_C$ is the heat flow to the storage (i.e., charging), $\dot{q}_D$ is the heat flow from the storage (i.e., discharging), $S_{cap}$ is the capacity of the heat storage, and $\overline{Q_{FLOW}}$ is the maximum allowable heat flow from and to the storage. In addition, $s$ is defined non-negative.

These formulations neglect the temporal variation of heat storage losses. Stationary heat loss depends upon the temperature difference between the storage and the environment, which is sometimes neglected in large-scale storage [69,75]. Henning and Palzer suggested calculating the stationary heat loss using storage tank parameters and the temperature difference between the storage and the environment [92]. Similarly, the dynamic losses can also be calculated if the pipe parameters and the temperature differences are known. However, for simplicity, both heat losses are often used as constant parameters, either as a percentage or assuming constant temperature differences over time. Salpakari et al. and Hedegaard et al. used such formulation with negligible losses to model TES in large-scale aggregated models [69,93]. The heat storage capacity $S_{cap}$ can be pre-calculated using formulations by Heinen et al., which consider specific heat capacity and density of water, temperature difference, and volume of the storage tank [67].

This study has not presented the modeling of buildings and their thermal characteristics. Rasku and Kiviluoma presented a lumped capacitance model used to describe the thermal dynamics of the detached housing stock [94]. The same approach could also be taken to represent stratification in TES and industrial heating processes.

New methods have been developed to design energy storage models and troubleshoot these systems in recent years. A review by Gao and Lu reveals that the typically used machine learning technologies applied in the research and development of energy storage systems include unsupervised learning, supervised learning (e.g., support vector machine/support vector regression), relative vector machine, decision tree/random forest, and regression algorithms), and deep learning (e.g., empirical model learning, artificial neural networks, convolution neural network, and recurrent neural network-based models) [95]. In addition, some machine learning technologies mix multiple algorithms and models to acquire maximum learning performance. The machine learning approaches are widely used for modeling TES with PCMs. For example, Tang et al. used a machine learning method to optimize a hybrid renewable energy system that incorporates PCM for active cooling applications [96]. Furthermore, Zhou et al. developed a generic method integrated with a single-layer feedforward neural network model and heuristic optimization algorithms to optimize a hybrid renewable system comprising PCM for five diverse climate regions [97]. The readers are suggested to read the comprehensive review by Gao and Lu to learn more about state-of-the-art machine learning methods for storage systems.

## Conclusion

This paper describes the principal P2H and TES technologies that are technologically innovative and economically viable and can improve energy efficiency and reduce $CO_2$ emissions. Based on the literature, we have sketched out P2H and TES technologies classifications. The most efficient and technologically matured P2H technologies for the European energy system are electric heat pumps, electric boilers, electric resistance heaters, and hybrid heating systems. Furthermore, the role of CHP in coupling power and heat sectors, especially in district heating, has been discussed. Among TES technologies, sensible and latent heat storages are mature and cost-effective technologies. However, other technologies such as thermo-chemical heat and thermo-mechanical heat storage are also prospective. Finally, the study presents the thirteen most promising technologies under these four TES categories.

Low-temperature electric heat pumps, electric resistance and electrode boilers, electric resistance heaters, and sensible and latent heat storage have a high TRL. Therefore, they can facilitate a large share of the heat demand. Besides low-temperature heat pumps for household heating, high-temperature heat pumps show promising potential in the food, paper, and chemical industries. Electric resistance boilers are suitable for low-temperature applications such as food and chemical industries. Electrode boilers are ideal for high-temperature process heating, such as steel and cement industries. Electric resistance heating for residential households is available as radiant heaters, baseboard heaters, wall heaters, underfloor heating systems, and electric furnaces. High-temperature industrial applications can use electric process heaters, besides conductive, inductive, high-frequency, magnetic direct current, and electrical infrared heating systems. Hybrid heating systems combining heat pumps with electric or gas boilers are also an essential solution. CHP plays a vital role in coupling power and heat sectors by supplying district and industrial process heating. An optimal combination of CHP with other P2H and TES technologies in district heating systems can facilitate flexible sector coupling of power and heat and shows excellent potential to increase renewable shares in the energy system. TES is considered an essential flexibility tool for smart energy systems to aid the smooth coupling of P2H and is expected to cover 60 % of total heat storage in 2050.

Heat pumps are modeled using COP formulations and taking the effect of temperature into account. Electric boilers and electric resistance heaters are modeled using similar equations but different power-to-heat ratios. Heat pump and electric boiler modeling consider the unit's operational status and power consumption limits. The electric boiler model avoids stratification effects. In the case of hybrid heating systems, each technology is modeled individually. Flexibility to such hybrid systems can be provided, either shifting the sources or shifting in time. CHP modeling formulations are presented for a generic case for backpressure and extraction turbines. Thermal energy storage is modeled using generic equations focusing on sensible hot water storage.

When the technology's performance depends on environmental temperatures, such as heat pumps and TES in insulated thermal mass, the effect of climate change is to alter the long-term weather averages and max/min values used for linear modeling of the technologies. Therefore, this feedback from the climate should be considered in large-scale energy system models.

In future research, it will be interesting to see the role of P2H in other geographical contexts. Another exciting area of study will be to analyze the combination of P2H, power-to-gas, and other options in an entirely sector-coupled (i.e., electricity, heat, and transport) energy system. Finally, further research should consider policy and regulatory behaviors in energy models to reflect on a more realistic analysis besides optimization models.

## Funding

This work received funding from TNO's internal R&D projects (No. 060.38253 and 060.42856), the ESTRAC project Integrated Energy System Analysis (No. 060.34020), European Union's Horizon 2020 research and innovation programme under the Marie Skłodowska-Curie grant agreement No. 765515 (ENSYSTRA), and the European Union's Horizon 2020 research and innovation programme under grant agreement No. 864276 (TradeRES).

## CRediT authorship contribution statement

**Md. Nasimul Islam Maruf:** Conceptualization, Methodology, Validation, Formal analysis, Investigation, Resources, Writing – original draft, Writing – review & editing. **Germán Morales-España:** Conceptualization, Methodology, Investigation, Writing – review & editing, Supervision. **Jos Sijm:** Conceptualization, Methodology, Investigation, Writing – review & editing, Supervision, Project administration, Funding acquisition. **Niina Helistö:** Investigation, Writing – review & editing. **Juha Kiviluoma:** Writing – review & editing.

**Table A1**
Technology readiness levels (TRL) and their basic properties. Adapted from [24].

| TRL | Description | Status |
|---|---|---|
| 1 | Basic principles observed and reported | Theory |
| 2 | Technology concept and/or application formulated | |
| 3 | Analytical and experimental critical function and/or characteristic proof of concept | Laboratory |
| 4 | Component and/or breadboard validation in laboratory environment | |
| 5 | Component and/or breadboard validation in relevant environment | |
| 6 | System/subsystem model or prototype demonstration in relevant environment | Prototype |
| 7 | System prototype demonstration in operational environment | |
| 8 | Actual system completed and qualified through test and demonstration | |
| 9 | Actual system proven through successful mission operations | Established |

## Declaration of Competing Interest

The authors declare that they have no known competing financial interests or personal relationships that could have appeared to influence the work reported in this paper.

## Acknowledgement

We want to express our gratitude to Prof. Dr. Olav Hohmeyer (Europa-Universität Flensburg) for his valuable and constructive suggestions during the planning and development of this research work.

## Appendix A

*Technology readiness level*

## Appendix B. Supplementary data

Supplementary data to this article can be found online at https://doi.org/10.1016/j.seta.2022.102553.